\documentclass{article}

\usepackage{arxiv}

\usepackage{bm}
\usepackage[english]{babel}
\usepackage[utf8]{inputenc} % allow utf-8 input
\usepackage[T1]{fontenc}    % use 8-bit T1 fonts
\usepackage{url}            % simple URL typesetting
\usepackage{booktabs}       % professional-quality tables
\usepackage{amsmath}       % math tool
\usepackage{amsfonts}       % blackboard math symbols
\usepackage{amssymb}
\usepackage{stackrel}
\usepackage{mathtools}  
\usepackage{nicefrac}       % compact symbols for 1/2, etc.
\usepackage{hyperref}
\usepackage[protrusion=true,
            expansion=true,final,babel]{microtype}

\usepackage{graphicx}
\usepackage{caption}
\usepackage[list=true]{subcaption}

\usepackage{siunitx}
\usepackage{tabularx}
\usepackage{placeins}
\usepackage{setspace}
\usepackage{fancyhdr}

\renewcommand{\vec}[1]{\mathbf{#1}}
\newcommand{\vers}[1]{\mathbf{\hat{#1}}}

\title{Sky Visibility Analysis for Astrophysical Data Return Maximization in HERMES Constellation}

\author{
  Andrea Colagrossi\thanks{Andrea Colagrossi, \href{mailto:andrea.colagrossi@polimi.it}{\tt andrea.colagrossi@polimi.it}}\\
  Postdoctoral Research Fellow\\
 Politecnico di Milano\\
 Via La Masa 34, 20156, Milano, Italy \\
\texttt{andrea.colagrossi@polimi.it} \\
   \And
 Jacopo Prinetto \\
 Ph.D. Candidate\\
 Politecnico di Milano\\
 Via La Masa 34, 20156, Milano, Italy \\
 \texttt{jacopo.prinetto@polimi.it}\\
    \And
    Stefano Silvestrini\\
    Ph.D. Candidate\\
 Politecnico di Milano\\
 Via La Masa 34, 20156, Milano, Italy \\
 \texttt{stefano.silvestrini@polimi.it} 
    \And
 Mich\`ele Lavagna\\
 Full Professor\\
 Politecnico di Milano\\
 Via La Masa 34, 20156, Milano, Italy \\
 \texttt{michelle.lavagna@polimi.it} 
}

\begin{document}
\thispagestyle{empty}
\large{\copyright 2020 Society of Photo-Optical Instrumentation Engineers (SPIE). \\ One print or electronic copy may be made for personal use only. Systematic reproduction and distribution, duplication of any material in this publication for a fee or for commercial purposes, and modification of the contents of the publication are prohibited.

\medskip

This document represents a preprint from \textbf{Andrea Colagrossi, Jacopo Prinetto, Stefano Silvestrini, Mich\`ele Lavagna, "Sky visibility analysis for astrophysical data return maximization in HERMES constellation," J. Astron. Telesc. Instrum. Syst. 6(4), 048001 (2020), \href{http://dx.doi.org/10.1117/1.JATIS.6.4.048001}{doi: 10.1117/1.JATIS.6.4.048001}."}

\medskip

Please check out the SPIE's official online version of this manuscript at \href{http://dx.doi.org/10.1117/1.JATIS.6.4.048001}{http://dx.doi.org/10.1117/1.JATIS.6.4.048001}.}

\clearpage

\setcounter{page}{1}
\maketitle

\begin{abstract}
HERMES is a scientific mission composed of 3U nano-satellites dedicated to the detection and localization of high-energy astrophysical transients, with a distributed space architecture to form a constellation in Earth orbits. The space segment hosts novel miniaturized detectors to probe the X-ray temporal emission of bright events, such as Gamma-Ray Bursts (GRBs), and the electromagnetic counterparts of Gravitational Wave Events (GWEs), playing a crucial role in future multi-messenger astrophysics.

During operations, at least three instruments, separated by a minimum distance shall observe a common area of the sky to perform a triangulation of the observed event. An effective detection by the nano-satellite payload is achieved by guaranteeing a beneficial orbital and pointing configuration of the constellation. The design has to cope with the limitations imposed by small space systems, such as the lack of on-board propulsion and the reduced systems budgets.

The paper describes the methodologies and the proposed strategies to overcome the mission limitations, while achieving a satisfactory constellation visibility of the sky throughout the mission duration. The mission design makes use of a high-fidelity orbit propagator, combined with an innovative mission analysis tool that estimates the scientific performances of the constellation. The influence of the natural relative motion, which is crucial to achieve an effective constellation configuration without on-board orbit control, is assessed. The presented methodology can be easily extended to any kind of distributed scientific space applications, as well as to constellations dedicated to Earth and planetary observation. In addition, the visibility tool is applicable in the context of the constellation flight dynamics operations, yielding optimized results and pointing plans based on actual satellite orbital positions.


\textbf{Full bibliographic reference:} \textit{Andrea Colagrossi, Jacopo Prinetto, Stefano Silvestrini, Mich\`ele Lavagna, "Sky visibility analysis for astrophysical data return maximization in HERMES constellation," J. Astron. Telesc. Instrum. Syst. 6(4), 048001 (2020), \href{http://dx.doi.org/10.1117/1.JATIS.6.4.048001}{doi: 10.1117/1.JATIS.6.4.048001}."}

\copyright 2020 Society of Photo-Optical Instrumentation Engineers (SPIE)[DOI: \href{http://dx.doi.org/10.1117/1.JATIS.6.4.048001}{10.1117/1.JATIS.6.4.048001}]

\end{abstract}

\section{Introduction}
%HERMES Introduction
HERMES is a scientific space mission dedicated to the detection and localization of high-energy astrophysical transients, such as Gamma-Ray Bursts (GRBs), and of the electromagnetic counterparts of Gravitational Wave Events (GWEs), whose detection probability is increased by distributing numerous sensors on Earth-bounded orbits. Therefore, a constellation of 3U nano-satellites is implemented to guarantee the distributed satellites architecture. 
During operations, at least three spacecraft, in the required configuration, shall observe GRBs events to perform triangulation and get a significant scientific content from the differential measurements \cite{HERMES_SWARM}. 

HERMES space segment is composed of a distributed architecture of 3U spacecraft, to fly on low Earth orbit. The science operations scenario is identified by means of analyses concerning the visibility of the sky and, thus, the expected number of GRBs in view of HERMES constellation. 
% In fact, the mission scenario selection is performed according to the estimates of the scientific performances: a key performance parameter measuring the sky visibility is defined, in order to evaluate the scientific return of the mission. 
The operational orbit and the injection strategy of the elements of the constellation determine the dynamics and the evolution of the relative orbits of the spacecraft. In addition, the pointing strategy is crucial to perform the alignment of the scientific instruments field of view (FOV). The FOV alignment, together with the mutual distances between the spacecraft (i.e. physical baseline), yields to the distance projected along the pointing direction of the scientific instruments (i.e. projected baseline). Furthermore, the effective GRBs detection by the nano-satellite payload is achieved by respecting numerous feasibility constraints on the orbital and pointing configuration of the spacecraft. 
The feasible projected baseline is evaluated to assess the fulfilment of the scientific objective, allowing to determine the region of the sky that can be triangulated and, hence, the expected number of triangulated GRBs. 

The current phase of the mission development is focused on the implementation of a technological pathfinder (TP) and a scientific pathfinder (SP) using a reduced constellation of 3+3 nano-satellites to be launched in 2021. HERMES-TP was selected by the Italian Ministry of University and Research (MUR), supported by the Italian Space Agency (ASI), to design, integrate and test the first 3 nano-satellites. HERMES-SP, funded by the EC-H2020, extended the constellation with a scientific pathfinder (SP) of additional 3 space elements. 
The options to launch the 6 satellites with a single launch or with 2 separate launches are still open, and the mission design had to cope with this variability.

%Paper Introduction
The paper describes the methodologies and the proposed strategies to cope with HERMES constellation, in the complete 6 spacecraft configuration. The limitations imposed by the small space system, such as the lack of on-board propulsion and the reduced systems budgets, shall be overcome, while achieving a satisfactory constellation visibility of the sky throughout the mission duration.
The proposed methodology combines a high-fidelity orbit propagator, including all the major environmental perturbations, with a sky visibility tool that analyzes the scientific constraints based on relative positions and pointing directions. 
The tool allows to perform the analysis of the preliminary constellation in any orbital scenario. Moreover, the investigation entails the possibility to compare different pointing strategies for the spacecraft, in order to select the nominal attitude motion that maximizes the visibility time. The presented tool is useful for flight dynamics operations planning, being flexible enough to be able to utilize in-flight data to generate optimal pointing schedules.
The influence of the orbital injection of the nano-satellites, both for a single and separate launch, is assessed in terms of the evolution of relative motion and achievable observation time. A propulsion system is not available; hence, it is critical to leverage natural motion and differential perturbations to achieve the desired configuration. 
The available results and the presented mission design work serve as a basis to implement an extended full constellation (FC) of $N$ satellites that will compose HERMES-FC.

%Motivation
This research work represents the basis for the design of satellite constellations dedicated to the observation of the sky, maximizing the coverage with numerous scientific detectors available. This is becoming particularly relevant for small satellites mission, whose advantage is based on the possibility to exploit a distributed system composed of many simple elements, rather than a unique scientific detector with advanced system capabilities. For this reason, the mission analysis shall exploit innovative techniques and tools capable to boost the constellation performances, while respecting the typical scientific requirements and the small system limitations.

%Sky generalization
The methods proposed in this paper are suitable for nano-satellites distributed scientific applications, being easily configurable to include any kind of scientific requirement that can be formulated as a mathematical relation. Moreover, the possibility to investigate the available scientific performances considering the natural relative dynamics is crucial for small space platforms without on-board propulsion. 

%Ground generalization
The developed tools can be further generalized to distributed missions dedicated to Earth and planetary remote sensing observation, being pointing direction and field of view of the scientific instruments completely configurable as per system and mission requirements. The proposed analytical optimization technique can easily include ground related scientific merit parameters, as long as they can be defined with mathematical sets and formulas, as will be described for the proposed HERMES applicative case in section \ref{sec:skyvis}. 

In summary, the intended scientific contribution of the paper can be summarized as:
\begin{itemize}
    \item to present the mission design for the HERMES constellation. In particular, being a nano-satellite mission, smart solutions are required to solve the shortcomings arising from limited mission resources (e.g. no orbital control, limited launch budgets, etc.)
    \item to describe the development of a sky visibility tool to assess scientific mission yield. The tool can be utilized by both constellation designer and operators to predict mission outcome or generate optimal pointing plans with respect to sky visibility performances; 
    \item to analyze different optimal pointing philosophies in terms of number of detected GRBs, sky regions coverage and triangulation accuracy; 
    \item to investigate the effect of number and frequency of spacecraft maneuvers required to change the payload pointing directions;    
    \item to investigate the robustness of the mission architecture against expected uncertainties and the possible extension of the proposed design, providing useful insights for future nano-satellite constellation designer.
\end{itemize}
%Paper Section description
In these regards, the paper is structured as follows.
Section \ref{sec:const_design} presents a discussion on orbital injection strategies, analyzing the main gravitational perturbations in low Earth orbit, to highlight the most effective way to achieve the desired inter satellite baseline, within the existing limitations in launch opportunity for small space systems.
Section \ref{sec:skyvis} describes the methods and the developed tool to analyze the sky visibility of the constellation, associated with certain instrument pointing directions. 
Section \ref{sec:op_point} discusses the effects of the definition and planning of the operational pointing directions on the astrophysical data return. In particular, the differences in GRB triangulation accuracy and in sky coverage are presented, as well as the influence of the pointing variations frequency. This section also presents the pointing directions optimization algorithm, designed to improve the scientific outcomes of the mission.
Section \ref{sec:Mission_Analysis_Results} presents the proposed HERMES constellation scenario. The robustness of the selected design strategies, with respect to mission uncertainties, is assessed with a statistical analysis. Different pointing strategies are compared to select the one providing the greater scientific results. Moreover, the guidelines to extend the constellation design to a larger full constellation are discussed.

\subsection{Scientific Requirements}
\label{sec:scireq}
HERMES payload is a simple but innovative detector \cite{FUSCHINO2019199} to probe the X-ray and Gamma-ray emission of bright high-energy transients. It is hosted in a single CubeSat Unit (1U) and the detector assembly is primarily composed of 60 scintillator crystals and 12 Silicon Drift Detectors (SDD) arrays, each with 10 independent cells. Both SDDs and crystals are employed to detect high energy
photons, the former in the X-ray range, the latter in the Gamma-ray range. The crystals are optically connected to the SDDs, and each crystal is read out by two SDD cells. The detectors are supported by a dedicated mechanical structure, a payload data handling unit, a power supply unit and an optical filter, to provide effective filtering of visible and UV-IR emissions. The payload is designed and assembled by the Italian National Institute of Astrophysics (INAF), which is also responsible for the payload operations.

HERMES mission design is strongly characterized by the payload operations, as well as the set of mission and scientific requirements. In particular, since a minimum number of GRBs shall be detected simultaneously by at least 3 space elements, the main scientific requirements affect the baseline between the satellites and the alignment of their FOVs. Formally, the requirements ask:
\begin{itemize}
    \item  at least 3 satellites shall have common FOV, within $\pm \ang{60}$ to maintain $50\%$ efficiency in the detector field;
    \item the physical baseline between at least 3 co-observing satellites shall be larger than $\SI{1000}{km}$;
    \item at least 2 couples of co-observing satellites shall have the baseline projected along the mean pointing direction of the scientific instruments larger than $\SI{1000}{km}$.
\end{itemize}
These requirements shall be satisfied within the HERMES system possibilities. In fact, HERMES satellites are equipped with complete attitude control subsystem, but no propulsion unit is present on-board. Therefore, the pointing direction of the payload and the FOV overlap can be controlled, but the physical baseline between the satellites shall be guaranteed by means of natural dynamics only. 

Additional scientific requirements led to constraints on the operational orbit of the constellation: detector lifetime is limited by the leakage current, which depends on the radiation environment, which is a function of orbit altitude and inclination. The requirements applied for the orbit design are:
\begin{itemize}
    \item the mission duration shall be greater than $\SI{2}{y}$;
    \item the altitude of the operational orbit shall be lower than $\SI{600}{km}$;
    \item the inclination of the operational orbit shall be lower than $\ang{30}$ or greater than $\ang{70}$.
\end{itemize}
The first requirement can be translated in a lower bound for the orbit altitude: a HERMES spacecraft, with a mass to area ratio in the order of $\sim \SI{50}{kg/m^2}$, has an orbital lifetime greater than $\SI{2}{y}$ for an initial orbit altitude greater than $\SI{450}{km}$, with average solar activity (i.e. $F10.7 \sim \SI{100}{sfu}$) \cite{pardini2001decay,PARDINI2006392}. Hence, the orbit design process is constrained to nearly-equatorial orbit or Sun Synchronous orbit (SSO) with $\SI{450}{km}\le h\le\SI{600}{km}$

\subsection{Orbital Simulator}
\label{sec:orb_prop}
The mission design relies on accurate modeling of the satellite orbital motion. In particular, the dynamical model shall be accurate enough to catch the natural relative dynamics between the spacecraft of HERMES constellation. The methods and the analyses presented in this paper are supported by an orbital propagator, developed and validated by the authors at Politecnico di Milano. The dynamical model takes into account all relevant perturbing forces acting in Low Earth Orbits, namely:
\begin{itemize}
    \item Earth geopotential based on EGM96 model. The accepted truncation error for the spherical harmonic order at the given altitude is selected according to the intrinsic accuracy of the EGM96 model, as per ECSS guidelines \cite{EGMNasa,ECSS-E-ST-10-04C}.
    \item Solar gravitational perturbation, with Sun’s position based on DE431 Ephemeris model \cite{folkner2014planetary}.
    \item Lunar gravitational perturbation, with Moon’s position based on DE431 Ephemeris model \cite{folkner2014planetary}.
    \item Atmospheric drag perturbation on flat surfaces. The atmosphere density is based on Jacchia-Roberts model. Solar activity coefficients (F107-AP index) are available from NOAA Solar flux prediction model - 2016 update \cite{jacchia1970new,roberts1971analytic,NOAA}.
    \item Solar Radiation Pressure (SRP) on flat surfaces. Solar activity is taken into account and the solar coefficients (F107-AP index) are available from NOAA Solar flux prediction model - 2016 update \cite{NOAA}.
\end{itemize}
The propagation is performed by using a Runge–Kutta–Fehlberg 7(8) method on a $C$ compiled code.

\section{Orbital Injection Strategies}
\label{sec:const_design}

The orbital injection of nano-satellites is crucial in determining the subsequent natural orbital evolution, since the lack of orbit control capabilities. In particular, the relative natural dynamics between the satellites of a constellation is specifically dependent from the orbital injection phase. In general, the relative positions of the spacecraft are not fixed in time. Thus, the constellation design has to optimize this dynamical evolution of the uncontrolled relative orbits, coping with differential environmental perturbations, orbit acquisition inaccuracies and limitation in the launch options availability. 

This aspect is particularly relevant for HERMES constellation design, which has to guarantee a minimum physical baseline between the satellites. The proposed solution for HERMES mission can be applied to a generic constellation of nano-satellites without on-board propulsion subsystem. The aim is to propose a smart solution to a typical nano-satellite mission limitation. In particular, this section presents a discussion on the solutions to naturally bound and passively control the relative motion of different nano-satellites injected in low Earth orbit. 
%In this scenario, the orbital injection of the different spacecraft has a big impact on the natural evolution of the constellation, together with the differential effect of the orbital perturbations. Therefore, the injection phase is crucial for the mission performances and it shall be taken into account in the design process, to identify the most suitable solution.

Two main options are possible to achieve a non-null relative distance between the satellites:
\begin{itemize}
    \item dedicated multiple injections - one per satellite - into different true anomalies at $t_0$;
    \item single injection of multiple spacecraft, with initial relative motion imposed by the deployer's release spring. In this case, it is convenient to release the satellites in groups of 3 elements (i.e. triplets), imposing a certain relative dynamics to each triplet.
\end{itemize}

The first option is highly sensitive to the release conditions: natural perturbations provoke a slightly controlled relative drift, which is emphasized by the launcher and deployer injection uncertainties. Moreover, this option asks for a dedicated launch, making it less convenient, especially for nano-satellites.

The second option can be realized even if no dedicated launch is settled. A triplet can be released in a single launch event with no dedicated maneuvers; $N$ satellites are released with $N/3$ single injections.  
The shift in true anomaly between different triplets' release events is settled at design level and it is affordable by either a single launcher maneuver or with separate launchers. 
Furthermore, the relative motion is actually imposed by the deployer spring authority, which overcomes the natural perturbations effects and the launcher injection uncertainties, as will be discussed in section~\ref{sec:robustness}. Hence, the expected scientific outcomes are more robust with respect to the release conditions.

% \subsection{Relative motion}
% \label{sec:rel_motion}
% The relative motion of the spacecraft has to guarantee the fulfilment of scientific requirements, especially in terms of relative distance between the elements of the constellation. In both injection options, the spacecraft are injected using dedicated deployment springs, which deliver a certain amount of $\Delta v$. 
A Formation-Flying-like triplet scenario would be beneficial to keep the spacecraft within acceptable bounded motion throughout the mission. Nevertheless, the lack of propulsion, the launcher inaccuracies and the spring injection $\Delta v$, prevent the bounded motion to be achieved. 

\subsection{Dedicated Multiple Injections} 
\label{sec:multiinj}
\begin{figure}[tb]
    \centering
    \begin{subfigure}[t]{0.4\textwidth}
        \centering
        \includegraphics[width=\textwidth]{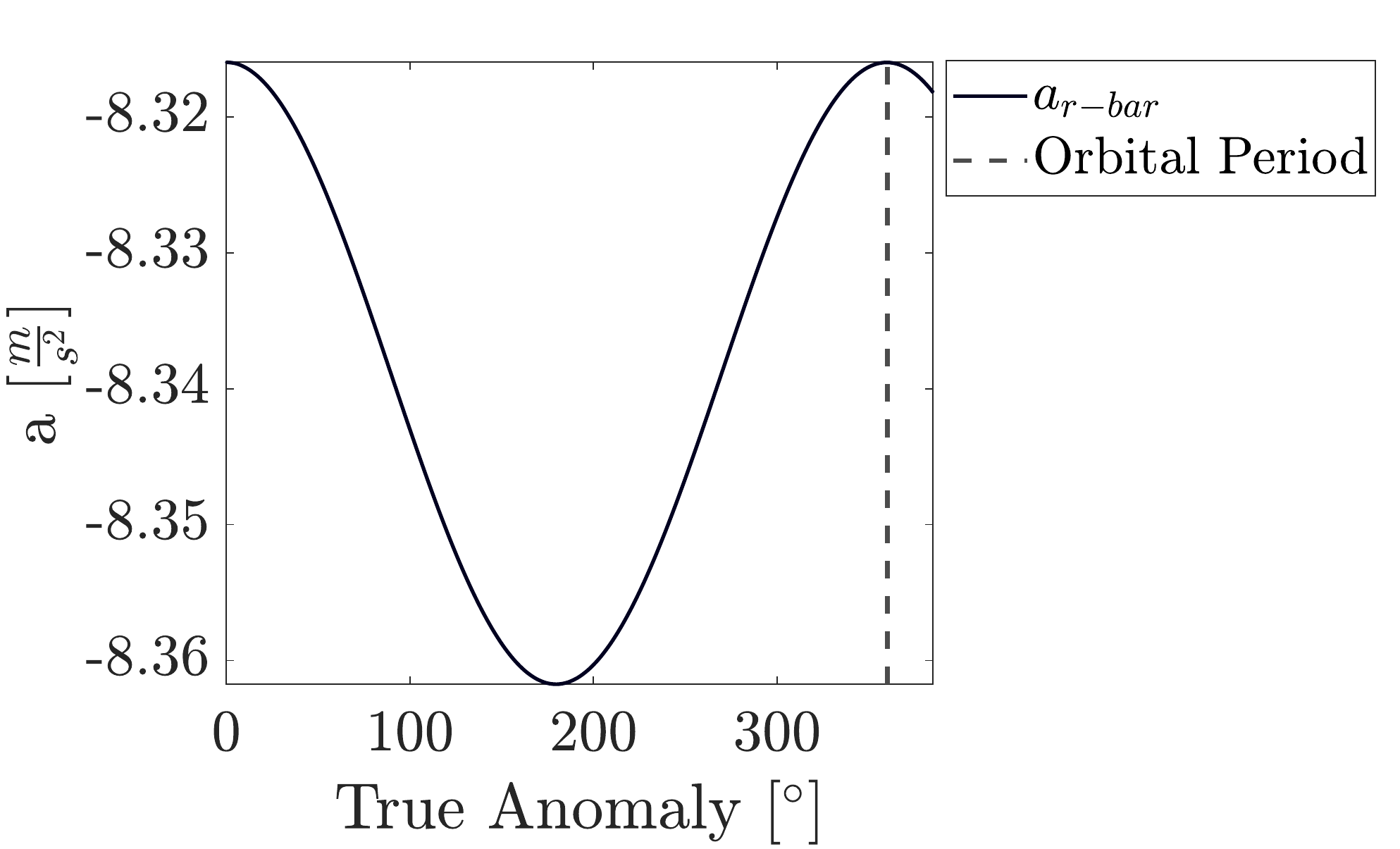}
        \caption{Radial direction component.}
        \label{fig:lvlh_acceleration_r}
    \end{subfigure}
    
    \begin{subfigure}[t]{0.4\textwidth}
        \centering
        \includegraphics[width=\textwidth]{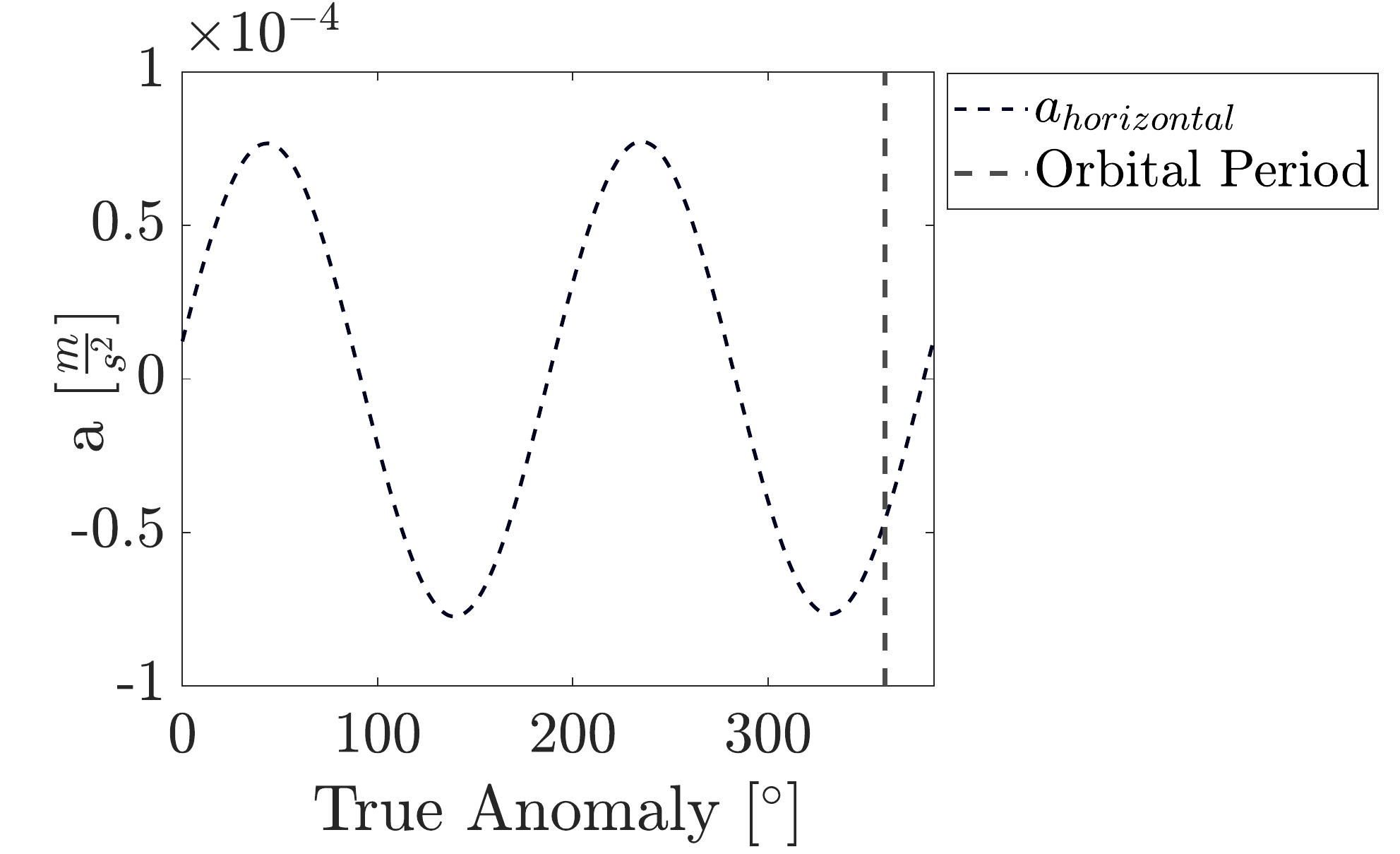}
        \caption{Horizontal direction component.}
        \label{fig:lvlh_acceleration_v}
    \end{subfigure}
    
    \begin{subfigure}[t]{0.4\textwidth}
        \centering
        \includegraphics[width=\textwidth]{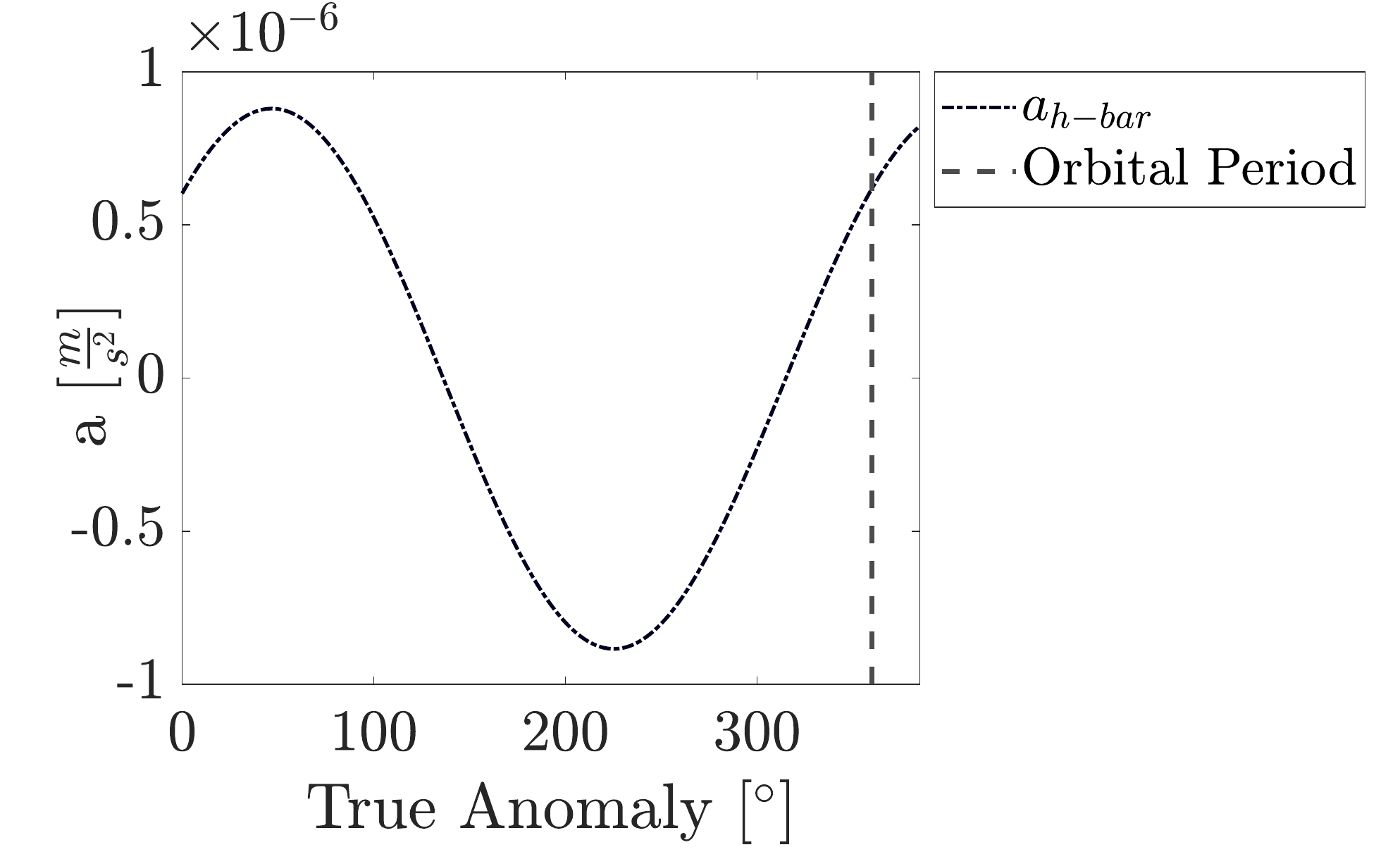}
        \caption{Orbital angular momentum direction component.}
        \label{fig:lvlh_acceleration_h}
    \end{subfigure}
    \caption{Orbital acceleration, with gravitational harmonics up to $J_{22}$, on a $\SI{550}{km}$ nearly-equatorial orbit, expressed in Local-Vertical/Local-Horizontal (\textbf{LVLH}) frame.}
    \label{fig:lvlh_acceleration}
\end{figure}

In the case of dedicated multiple injections, the initial true anomaly separation yields differential perturbation acceleration resulting in a secular drift of the satellites. Formation flying and, in general, distributed space systems orbital design often relies on linearized, unperturbed dynamical models given the short relative distances acquired and maintained during the mission. HERMES spacecraft are rather peculiar because of the large distances involved, which are reached during the mission, coupled with the incapability to perform orbit control. The relative motion is hence unbounded, but it is quasi-periodic. Given the importance of their relative motion, it is critical to assess the influence of perturbations, initial true anomaly and spring injection, on the long-term orbital motion.  
In case Keplerian motion is considered, the relative positions of the satellites are easily predicted throughout the whole mission. In addition, the relative positions are invariant to the initial true anomaly separation. Obviously, in reality, this is not the case due to the irregular shape of the Earth, leading to differential perturbations. Natural motion due to Spherical Harmonics is highly dependent on the initial true anomaly of the three satellites.  Figure~\ref{fig:lvlh_acceleration} shows the orbital acceleration, as a function of the satellite true anomaly, with gravitational perturbations up to $J_{22}$, along a $\SI{550}{km}$ nearly-equatorial orbit. 

The acceleration components are represented in the Local-Vertical/Local-Horizontal (LVLH) frame, which is defined by the radial vector pointing outward (i.e. zenith direction) and by the orbital angular momentum. The third axis completes the right-handed triad, in the horizontal direction with respect to the ground.  The latter component is oriented in a direction very close to the along-track (i.e. velocity) direction. Note that, for circular orbits, the horizontal component is parallel to the velocity direction.

A differential value in the along-track acceleration component, shown in figure~\ref{fig:lvlh_acceleration_v}, builds up a non-null relative velocity, which yields disparity in the accumulated orbital motion.  This differential effect is highly dependent on the initial true anomaly of the satellites, resulting in a non-static relative motion that is hardly predictable during the design phase, because of inherent uncertainties on $\theta(t_0)$. In fact, true anomaly precise injection is hardly achievable in typical nano-satellite launch and deployment phases.
Moreover, it may be noted how the periodicity of the perturbation acceleration in the horizontal direction is larger than $2\pi$ (see figure~\ref{fig:lvlh_acceleration_v}). This is due to the physical source of the perturbation, which is the Earth geopotential. Hence, the period of the resultant acceleration is the revisit time of Earth location, rather than orbital period with respect to the inertial frame. Indeed, it is possible to quantify the angular displacement, after one orbit, of these two reference frames as $\sim 24^{\circ}$, for the considered $\SI{550}{km}$ nearly-equatorial orbit.

\begin{figure}[t]
    \centering
    \begin{subfigure}[t]{0.4\textwidth}
        \centering
        \includegraphics[width=\textwidth]{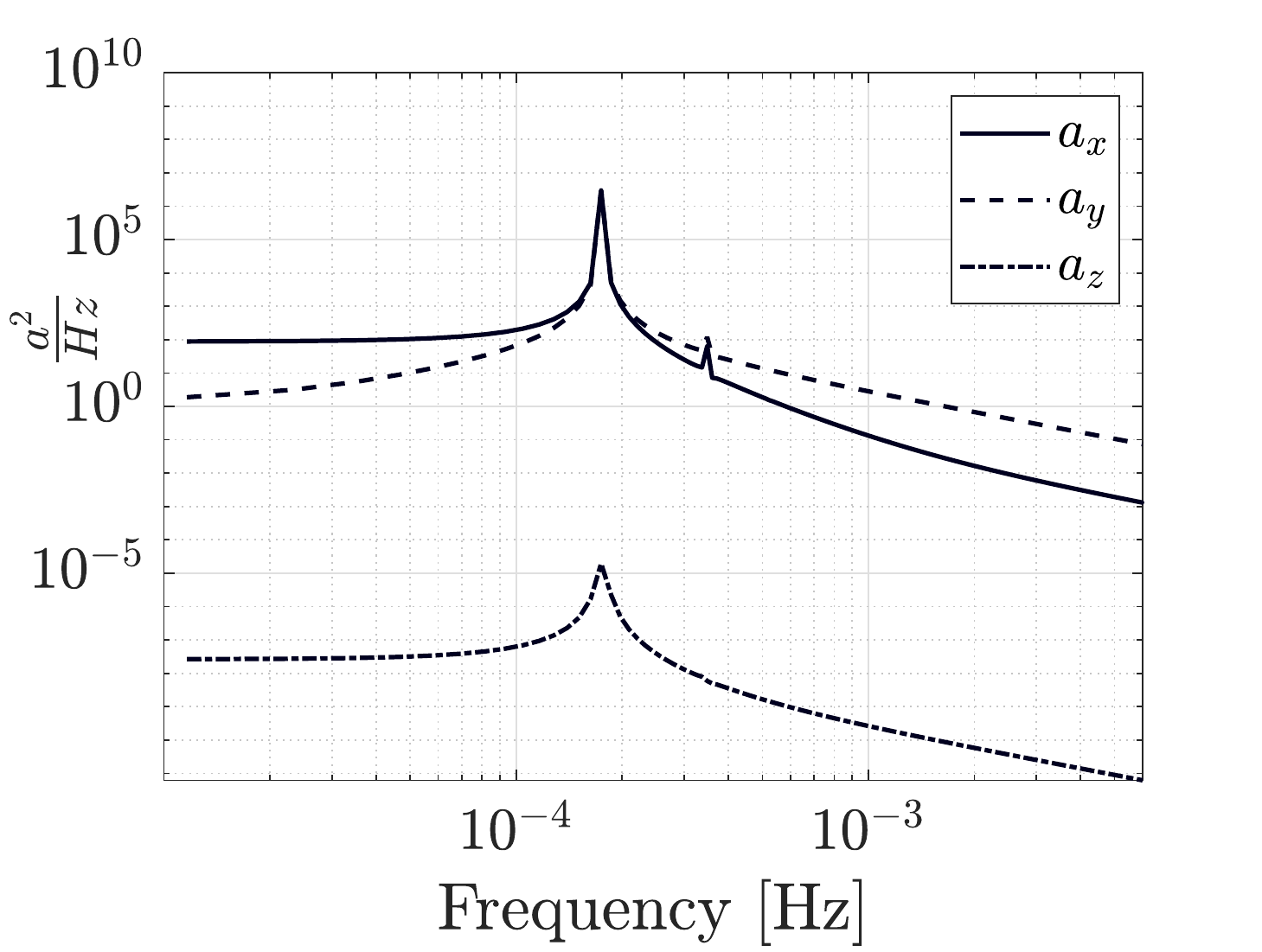}
        \caption{Keplerian acceleration.}
        \label{fig:psd_acceleration_kep}
    \end{subfigure}

    \begin{subfigure}[t]{0.4\textwidth}
        \centering
        \includegraphics[width=\textwidth]{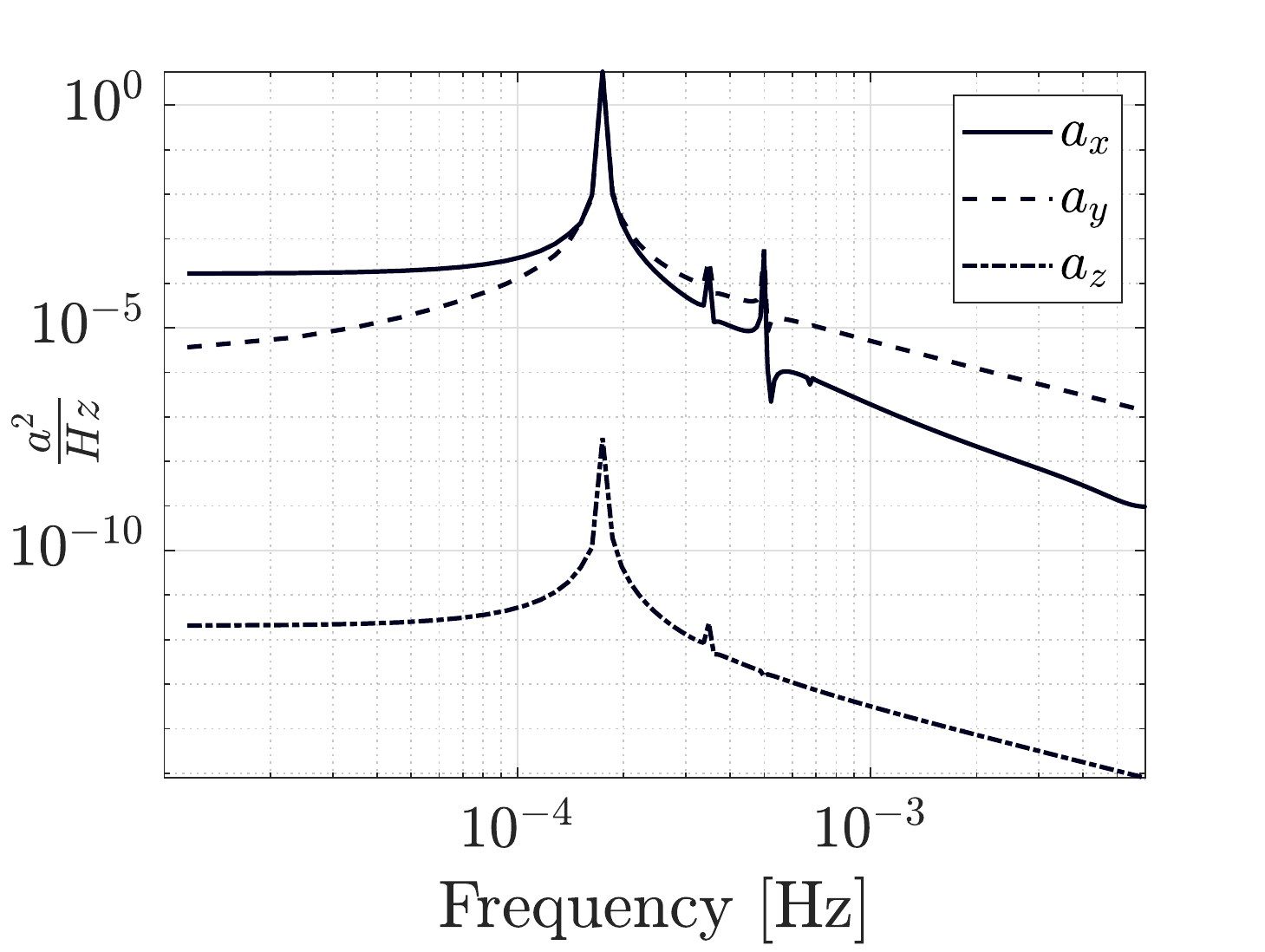}
        \caption{Gravitational perturbation acceleration up to $J_{22}$.}
        \label{fig:psd_acceleration_pert}
    \end{subfigure}

     \caption{Power Spectral Density (PSD) of orbital acceleration on a $\SI{550}{km}$ nearly-equatorial orbit, expressed in \textbf{inertial frame}. The PSD frequency range is shown within the validity of Nyquist-Shannon sampling theorem, with reference to the data-set sample time (i.e. $\SI{60}{s}$).}
    \label{fig:psd_acceleration}
\end{figure}

The perturbation accelerations are several orders of magnitude lower than the dominant Keplerian dynamics, as shown in figure~\ref{fig:psd_acceleration}. The Keplerian acceleration power spectral density (PSD), reported in figure~\ref{fig:psd_acceleration_kep}, refers to the osculating orbit point-wise two-body acceleration. Being osculating orbits, beside the frequency peak corresponding to the orbital motion, the PSD shows a small peak at twice the frequency, which is linked to the periodicity of magnitude of the radius of the osculating orbital motion.  The gravitational perturbation acceleration PSD, in figure~\ref{fig:psd_acceleration_pert}, shows the peaks corresponding to the orbital mean motion and its integer multiple due to the spherical harmonic terms. 
The accumulated differential velocity with initial true anomaly separation of $\SI{20}{\degree}$ is in the order of $\si{5-10}{m/s}$, according to the acceleration field analysis up to $J_{22}$. 

This result prevents the multiple dedicated injection from being a good option, beside the fact of increasing the cost for dedicated launches or launcher phasing maneuvers. Even for multiple dedicated launches, the nano-satellites deployment is performed using a pre-loaded spring, which injects the spacecraft into orbit. In order to inject a spacecraft at a defined true anomaly (e.g. the case of multiple injections), the natural motion is exploited. In particular, a harmonic motion cross-track is pursued to keep the satellite around a specific true anomaly, vanishing the effects of the initial spring $\Delta v$. In this case, gravitational perturbations drive the natural motion. This technique is valid for any nano-satellite injection. Indeed, since the spring release impulse is inevitable, one smart solution to maintain a given true anomaly without drifting is to be injected across-track.

% \begin{figure}[t]
%     \centering
%     \includegraphics[scale=0.6]{aLVLH-eps-converted-to.pdf}
%     \caption{Orbital acceleration expressed in the \textbf{LVLH frame}. The motion is propagated using only gravitational harmonics up to $J_{22}$, in a $\SI{550}{km}$ nearly-equatorial orbit.}
%     \label{fig:lvlh_acceleration}
% \end{figure}

% \begin{figure}[t]
%     \centering
%     \includegraphics[scale=0.6]{psd-eps-converted-to.pdf}
%     \caption{Power Spectral Density plot of orbital acceleration in the \textbf{inertial frame} up to $J_{22}$, in a $\SI{550}{km}$ nearly-equatorial orbit. The PSD frequency range is shown within the validity of Nyquist-Shannon sampling theorem, with reference to the integrator step time.}
%     \label{fig:psd_acceleration}
% \end{figure}

\subsection{Single Injection of Multiple Spacecraft} 
\label{sec:singleinj}
\begin{figure}
    \centering
    \includegraphics[width=0.6\textwidth]{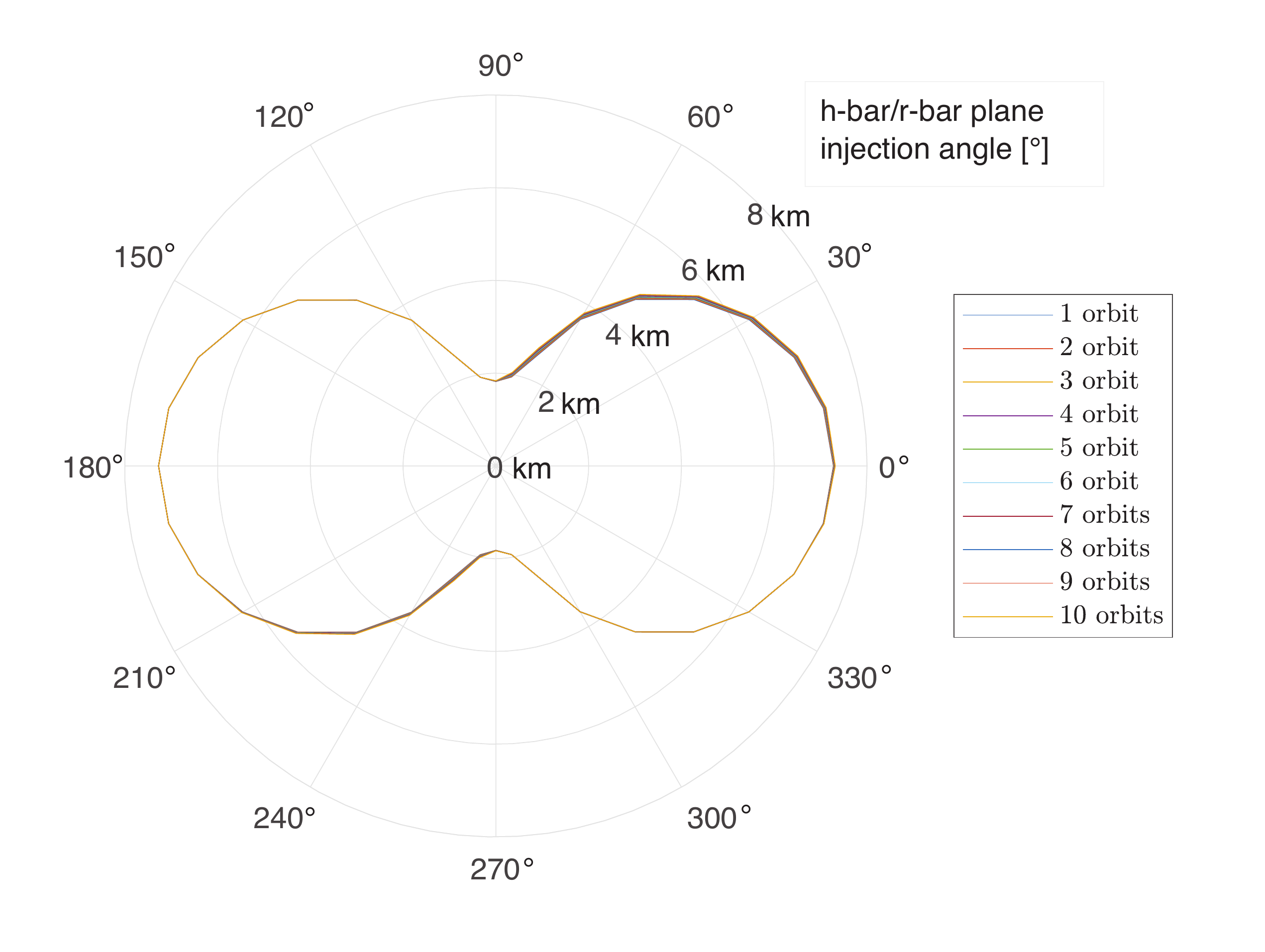}
    \caption{Maximum achievable distance during bounded cross-radial motion as a function of the injection angle with a spring $\Delta v = 2\frac{m}{s}$.}
    \label{fig:bounded_rel_orbits}
    \end{figure}

\begin{figure}
    \centering
    \includegraphics[width=0.6\textwidth]{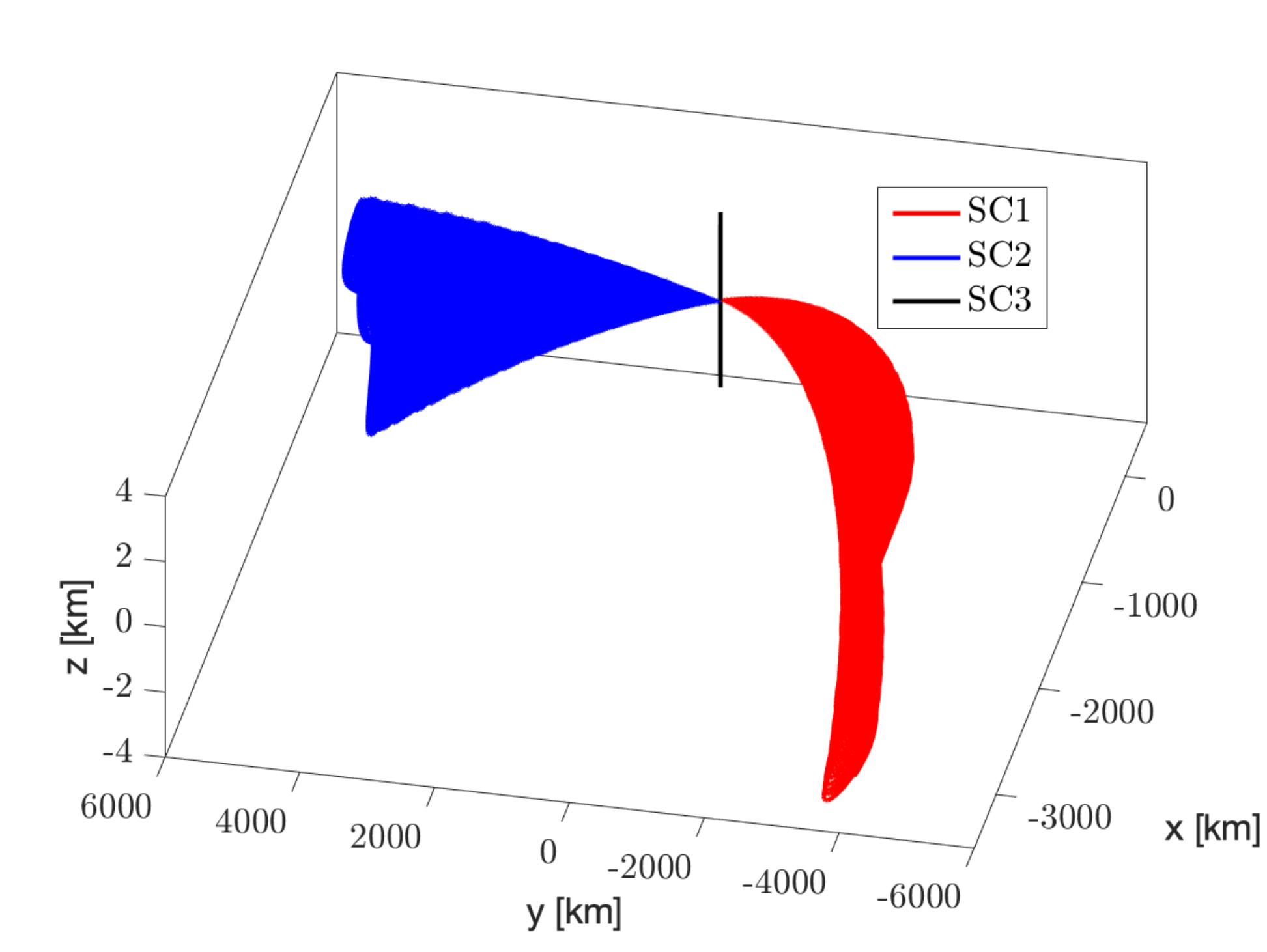}
    \caption{Injection strategies forcing relative drift to quickly achieve the required relative distance. The satellites are injected along $+v,\ -v,\ h$-bar with a spring of $\Delta v = 2\frac{m}{s}$. The physical baseline is achieved in $\sim\SI{3}{d}$.}
    \label{fig:rel_motion}
\end{figure}

In the case of single injection, the relative motion could be restrained to bounded relative orbits only if along-track injection $\Delta v$ is null. Such condition would be beneficial with respect to the temporal evolution of the relative distances between the satellites, but several considerations need to be made if a single injection strategy is analyzed. 
As said, a bounded relative motion can be achieved by vanishing the along-track differential velocity. Such method relies on the \textit{energy-matching} principle in which the energy of the absolute orbits of the different spacecraft are equal \cite{alfriend2009spacecraft,scharf_ff_survey}. An impulse on the radial-cross plane results in a bounded motion. Nevertheless, the entity of the $\Delta v$ provided by the spring is not enough to inject the spacecraft into three different orbits that result in a bounded relative motion respecting the minimum required mutual distance.
Figure~\ref{fig:bounded_rel_orbits} shows the maximum achievable distance during bounded cross-radial motion as a function of the injection angle in the radial-cross plane. The maximum reachable relative distance is in the order of $10^1\ km$, which is far from being acceptable for the imposed scientific requirement of $10^3\ km$. This sets the need to introduce a relative unbounded drift between the satellites to achieve, at least for a certain amount of time, the required baseline. Indeed, inserting a relative drift makes the formation loose with a periodic relative motion. The satellites need to be injected into different directions to prevent two, or more, satellites from being too close during mission operations. A satisfactory strategy consists in injecting the three satellites along $+v,\ -v,\ h$-bar respectively; the injection condition is expressed in the LVLH frame of the launcher at injection time. The satellites injected along-track start drifting, whereas the spacecraft injected cross-track evolves in  a harmonic motion around the injection point. In this condition, the resultant relative motion propagated for $\sim\SI{3}{d}$ is depicted in figure~\ref{fig:rel_motion}.

In a realistic scenario, a true simultaneous injection is hardly feasible. However, the above considerations have been proven to be still valid when dealing with short-delayed injections ($\sim 10/15$ min). The nano-satellites deployment is performed also in this situation using a pre-loaded spring, which injects the spacecraft into orbit. The analysis on the long-term perturbed propagation has shown that spring release injections with at least $\SI{0.25}{m/s}$ dominate the relative dynamics imposed by the natural perturbations, as presented in section~\ref{sec:Mission_Analysis_Results}. Hence, the design can be done regardless of the initial injection true anomaly, as will be discussed in the followings.

\section{Methods for Sky Visibility Evaluation}
\label{sec:skyvis}

The HERMES constellation sky visibility depends on the pointing direction of the satellite's FOV. In fact, despite the orbital control is not available and the positions of the spacecraft are completely dependent from the imposed natural motion, the attitude control is accessible to align the scientific instruments on different regions of the sky, allowing the overlapping of at least 3 satellites' FOVs. In particular, the accurate selection, alignment and maintenance of certain pointing directions can be exploited to enhance the whole mission scientific return, by increasing the time and the area of the sky that can be used to perform GRBs triangulation.

%In particular, the amount of observed sky area times the triangulation time is reflected in the number of detectable GRBs and, thus, it results in the performance parameter of the mission design. 

% The output of the mission design is selected according to the estimates of the scientific performances, which are evaluated according to the performance index of sky visibility achieved by the constellation. The performance index takes into account the area of the sky that is visible and triangulable, while satisfying all imposed scientific constraints and requirements; the triangulable sky area availability in time is part of the performance index as well. 

The resulting performance index for the constellation is defined as an area of the sky that can be viewed by at least three satellites for a certain amount of time and, thus, is denoted as time-area parameter: $tA$, whose dimensional units are steradians-day $\;[\si{\steradian-\day}]$.
Time-area parameter is defined so to be well suited to estimate the number of expected GRBs detected by the HERMES constellation. In fact, according to previous missions the rate of measurable GRBs is known. For example, FERMI-GBM \cite{meegan2009fermi,gruber2014fermi} mission detects $\sim\SI{250}{GRB}$ per year in the whole sky not covered by the Earth (i.e. $\sim\SI{8}{\steradian}$), which corresponds to almost 1 GRB every $\SI{1.5}{d}$ in the FOV of $\sim\SI{8}{\steradian}$, or $\SI{0.083}{GRB / \steradian-\day}$. 

\subsection{Sky Visibility Tool}
\label{sec:Sky_Visibility_Tool}

The evaluation of the performance time-area parameter is performed with a sky visibility tool developed and validated by the authors at Politecnico di Milano. 

The visibility tool takes as input the trajectory parameters of the $N$ satellites, it propagates the orbital dynamics with the orbital propagator described in section \ref{sec:orb_prop} and evaluates the time-area performance parameter. The developed tool, which is extensively used for HERMES mission analysis, is actually a generic framework that can provide useful insights for any mission that includes visibility optimization and pointing constraints. In addition, it is foreseen to be employed during operations planning. The optimized pointing can be predicted in the short time horizon using in-flight orbital data.

\begin{figure}[tb]
    \centering
    \includegraphics[width=0.5\textwidth]{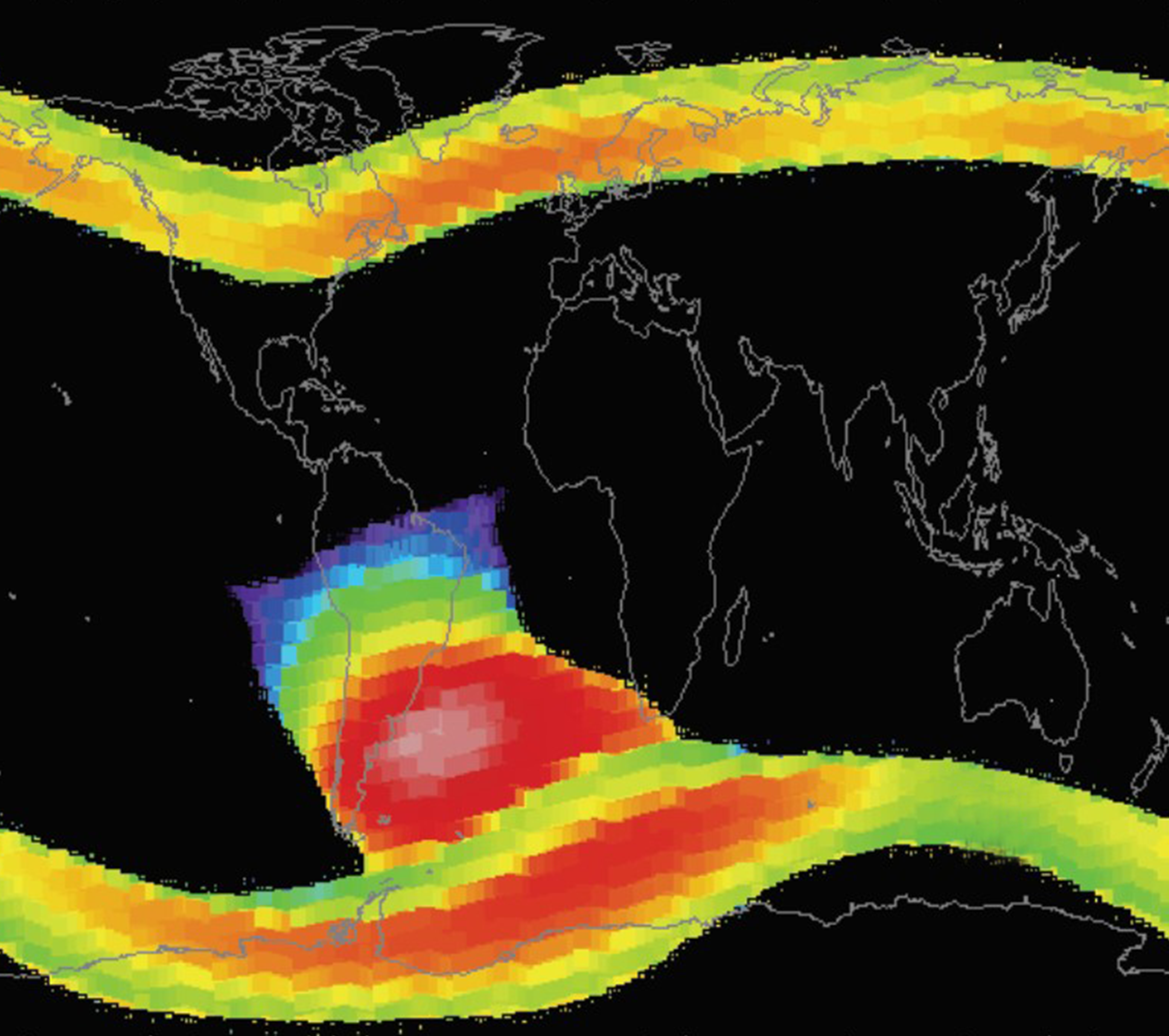}
    \caption{Large radiation flux regions. Allowed positions in black.}
    \label{fig:earth_flux}
\end{figure}

\paragraph{Satellite Position Constraints}
The software implements the scientific requirement and constraints, as well as possible system and platform constraints. At each iteration steps, the tool verifies if the satellites position with respect to the Earth surface is compatible with the observations. For example, large radiation flux regions are not compatible with scientific instrument operations, as high leakage currents in the active detectors may damage the payload. Similarly, when the satellites are in eclipse regions or ground stations areas, they may be not able to correctly observe the sky for power or antenna orientation constraints. For this reason, position requirements are enforced at all times for any $i-$th satellite $S_i$ as:
\begin{equation}
    S_i \in \mathcal{A} \Longleftrightarrow \vec{r}_i\in\mathcal{P},
    \label{eq:position_req}
\end{equation}
where $\mathcal{A}$ is the set of active satellites, $\vec{r}_i$ is the position vector of $S_i$ and $\mathcal{P}$ is the set of allowed positions. 
% For example, the set of allowed positions in the case large radiation flux regions are not compatible with scientific instrument operations is reported in black in figure~\ref{fig:earth_flux}.
For example, figure~\ref{fig:earth_flux} shows the allowable regions in black with respect to the radiation flux constraint.
In general, any constraints on the scientific operations of a nano-satellite dependent from the orbital position can be implemented in a similar way.

\begin{figure}[tb]
    \centering
    \begin{subfigure}[t]{0.75\textwidth}
    \centering
    \includegraphics[width=\textwidth]{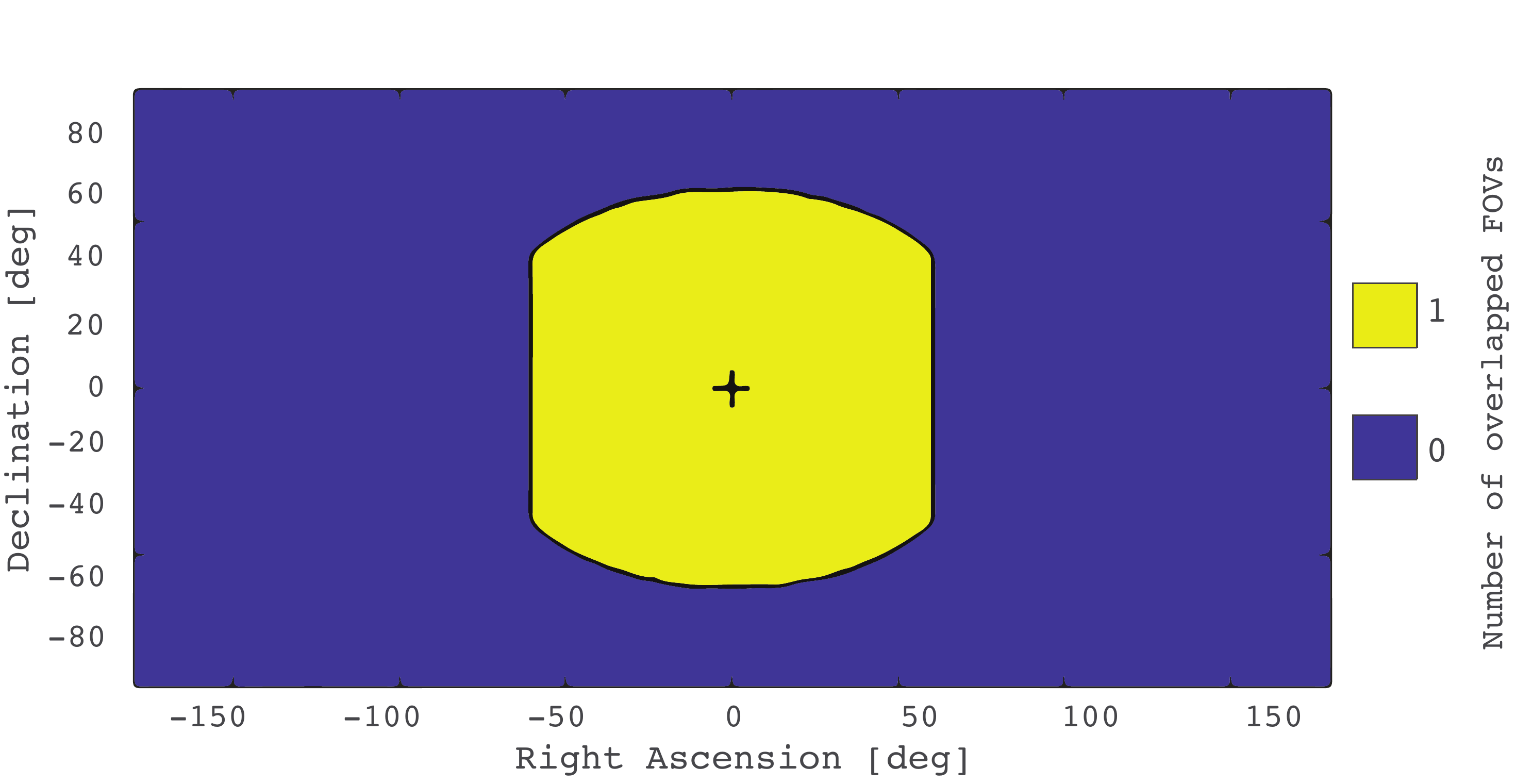}
    \caption{2D projection.}
    \end{subfigure}
    
    \begin{subfigure}[t]{0.6\textwidth}
    \centering
    \includegraphics[width=\textwidth]{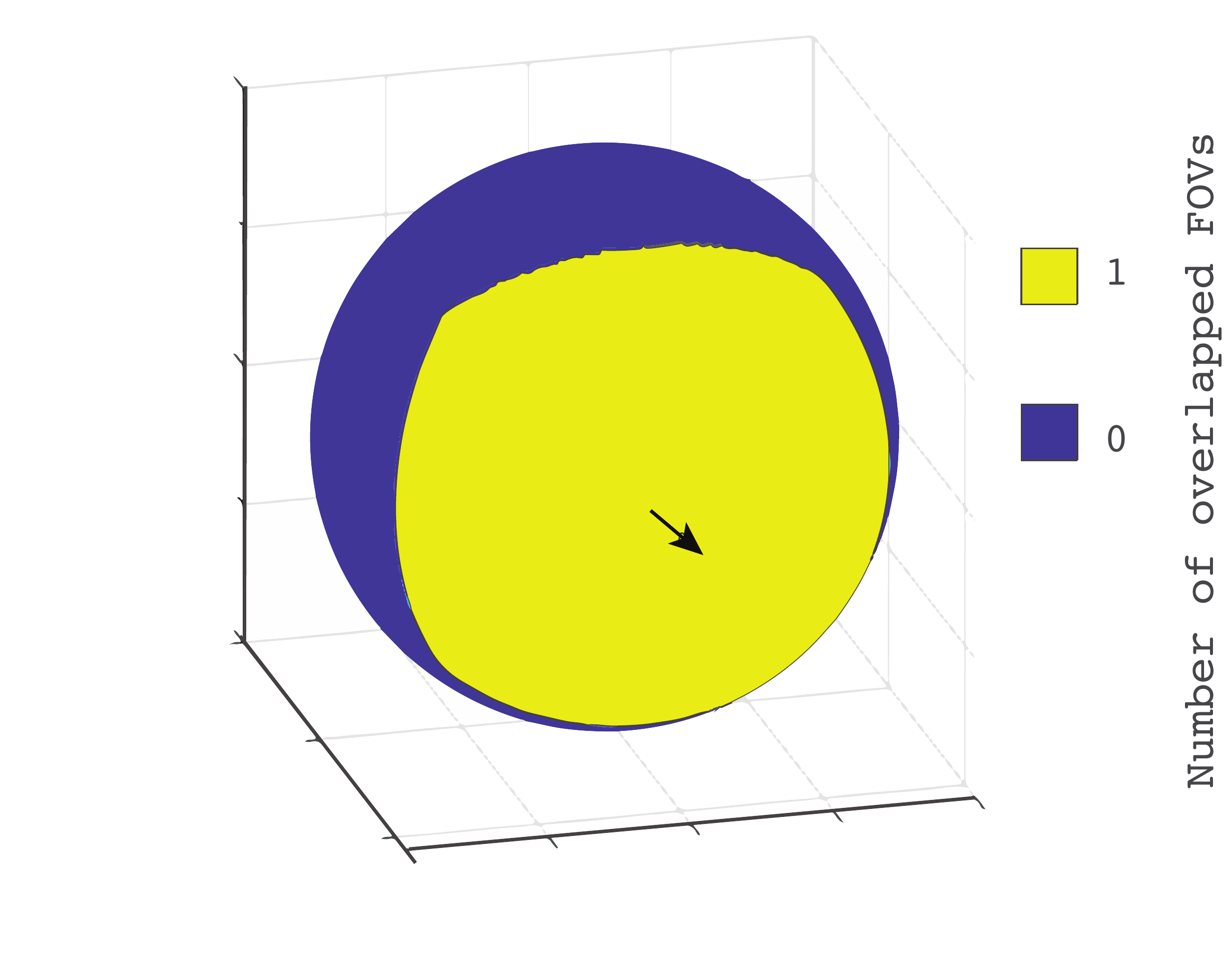}
    \caption{3D view.}
    \end{subfigure}
    \caption{Single instrument FOV. Pointing direction, $\vers{p}$, in $RA=\SI{0}{deg}$ and $DEC=\SI{0}{deg}$.}
    \label{fig:single_FoV}
\end{figure}

\paragraph{Satellite FOV}
The FOV of the satellite is characterized by the specific instrument installed on-board. It can be defined as a region of the sky that is enclosed within certain geometrical boundaries. In HERMES, the instrument's FOV is the semi-sphere in front of the detector plane. However, the sensitivity of the detector has a cosine profile with respect to the line of sight (LOS) and, thus, the full width at half maximum (FWHM) of the instrument is $\SI{120}{deg}$. As a consequence, for each active satellite the effective FOV is defined as the portion of the celestial sphere with an angular distance lower than $\SI{60}{deg}$ with respect to the instrument axial planes, $\Pi_x$ and $\Pi_y$, as reported in figure~\ref{fig:single_FoV}. In particular, for a given pointing direction of the line of sight (LOS) of the instrument, $\vers{p}$, which is found by the intersection of the two axial planes as:
\begin{equation}
   \vers{p} =  \Pi_x \cap \Pi_y,
\end{equation}
the FOV of the $i-$th satellite, $\mathcal{F}_i$, is defined 
as:
\begin{equation}
    \mathcal{F}_i=\left\{\vers{s} \;|\; \left(\vers{s}\angle\Pi_x \le \frac{\pi}{3}\right) \; \wedge \; \left(\vers{s}\angle\Pi_y \le \frac{\pi}{3} \right)  \right\},
    \label{eq:FOV}
\end{equation}
where $\vers{s}$ is a generic direction in the sky.
    
\paragraph{Satellite Pointing Constraints}
The developed tool is capable to include pointing constraints, in order to accommodate possible scientific requirements penalizing certain regions of the sky where the scientific performances are not acceptable. 
% They are soft pointing constraints, resulting in just not taking into account the associated measurements to estimate the scientific results. The hard pointing constraints, associated to regions of the sky that shall not be pointed, are assumed to be always respected, thanks to the attitude control system of the nano-satellite \cite{colaIAC2019}. 
The pointing constraints are divided in \textit{soft} and \textit{hard} depending on the consequences of constraint violations. 
On one hand, the soft pointing constraints identify regions of the sky where the satellite could, in principle, be pointing but without yielding valuable scientific measurements (e.g. high instrument noise). On the other hand, hard pointing constraints identify forbidden regions to be pointed, due to the potential degradation of the scientific instrument if not respected. Both constraints are assumed to be always respected, thanks to the attitude control system of the nano-satellite \cite{colaIAC2019}.
The soft pointing constraint region is formalized for a given constraint direction, $\vers{c}$, as:
\begin{equation}
    \mathcal{C}=\left\{\vers{s} \;|\; \cos^{-1}\left( \vers{s}\cdot\vers{c} \right) \le \alpha_c \right\},
\end{equation}
where $\alpha_c$ is the exclusion angle.
Then, the effective $i-$th FOV simply results from:
\begin{equation}
    \mathcal{F}_{E_i}=\mathcal{F}_i\;/\;\mathcal{F}_i\cap\mathcal{C}.
\end{equation}
It shall be noted that only active satellites belonging to $\mathcal{A}$ have an effective FOV. For satellites that are not active it results:
\begin{equation}
    \mathcal{F}_{E_i}|(S_i \notin \mathcal{A})=\emptyset.
\end{equation}

\paragraph{Triangulable Field of View} 
According to the imposed scientific requirements, a region of the sky can be triangulated if it is contained in the FOV of at least three satellites. Hence, the overlap of at least three effective FOVs is defined a $j-$th triangulable FOV as:
\begin{equation}
    \mathcal{F}_{T_j}=\mathcal{F}_{E_1}\cap\mathcal{F}_{E_2}\dots\cap\mathcal{F}_{E_i}\dots\cap\mathcal{F}_{E_N} \text{ with } N\ge3.
\end{equation}

\begin{figure}[tb]
    \centering
    \includegraphics[width=0.75\textwidth]{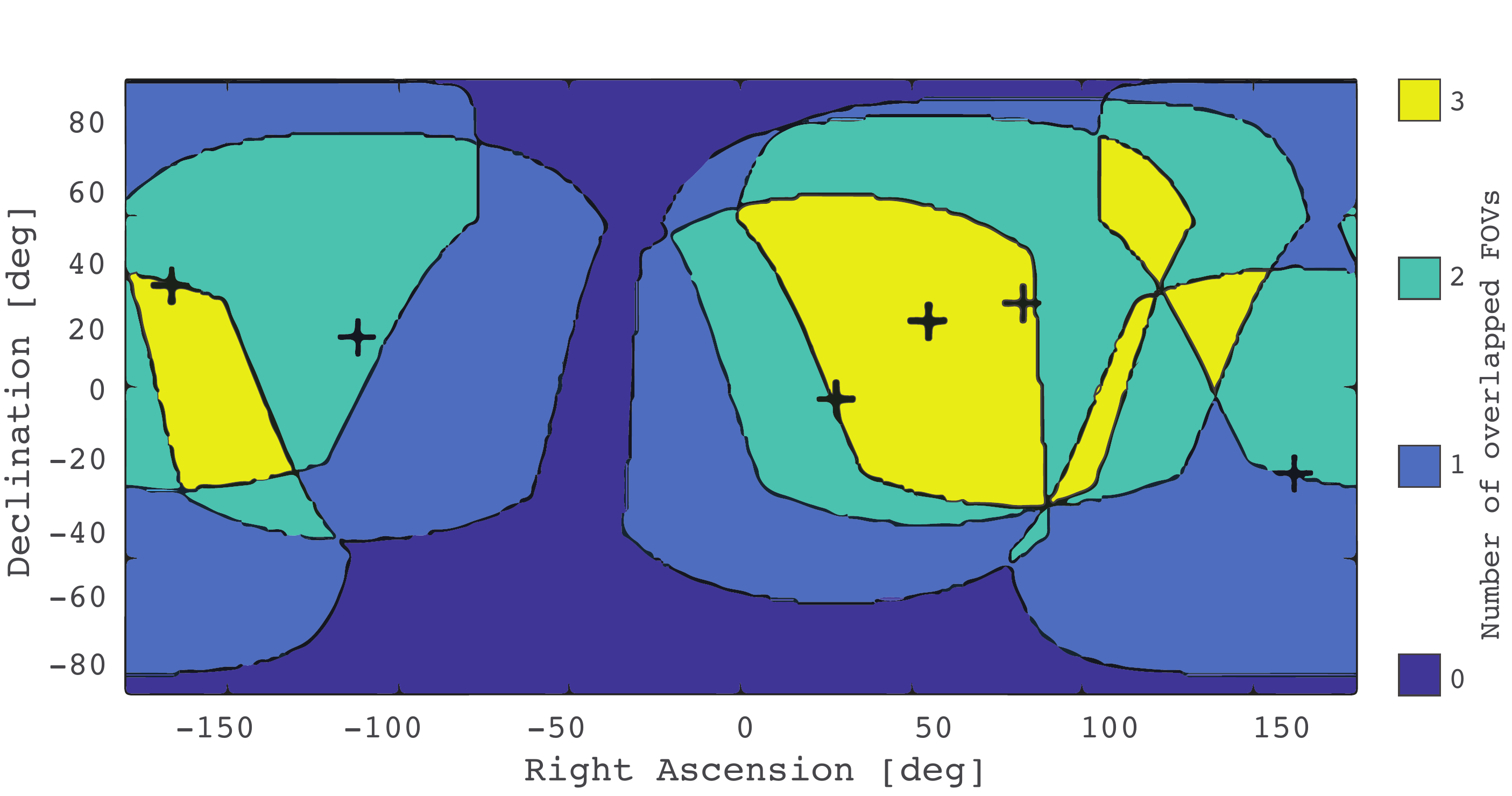}
    \caption{Satellite FOVs overlap.}
    \label{fig:FoVs_overlapping}
\end{figure}

Figure~\ref{fig:FoVs_overlapping} shows an overlapping map in which six satellites are active at a given epoch. The triangulable FOVs are the yellow regions in the map.
 
\paragraph{Baseline Requirements} 
The physical and the projected baselines for all the possible triangulable regions are evaluated. Indeed, for any $j-$th triangulable FOV there exist at least three co-observing active satellites, whose physical baselines are immediately available from their orbital position. The projected baselines are computed knowing the pointing directions of each element.
If the baseline requirements, described in section \ref{sec:scireq}, are satisfied, the triangulable FOV region is kept in the computation. Otherwise, the triangulable region is cancelled and set equal to zero.

\paragraph{Constellation Field of View}
 
\begin{figure}[tb]
    \centering
    \includegraphics[width=0.75\textwidth]{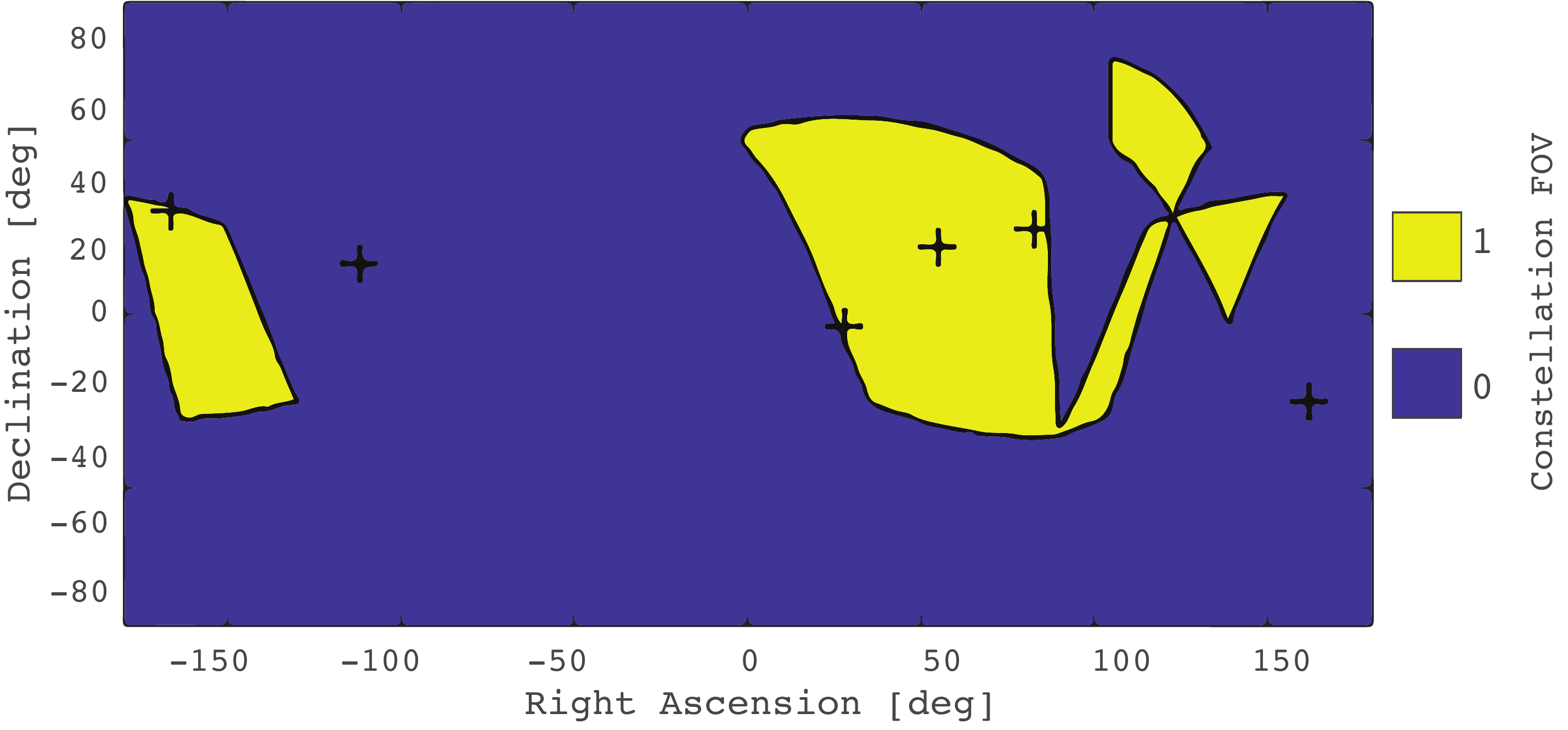}
    \caption{Constellation FOV.}
    \label{fig:triangulable_regions}
\end{figure}

At any epoch, each point in the sky visible by at least one active triplet (i.e. $\mathcal{F}_{T_j}\neq\emptyset$) is considered as observable and triangulable; the total area observed at any time is computed by integrating the observable points over the whole celestial sphere and it is denoted as constellation FOV. 
The constellation FOV results from the union of all the triangulable FOVs respecting the baseline requirements as:
\begin{equation}
    \mathcal{F}_{C}=\mathcal{F}_{T_1}\cup\mathcal{F}_{T_2}\dots\cup\mathcal{F}_{T_i}\dots\cup\mathcal{F}_{T_M} \text{ with } M\ge1.
\end{equation}
The constellation FOV area is measured in steradians [sr]. Figure~\ref{fig:triangulable_regions} shows the constellation FOV for the considered example. 
    
\paragraph{Time-area parameter and estimate of observable GRBs.}
    
The instantaneous observable area (i.e. the constellation FOV) is integrated over the whole mission lifetime to obtain the time-area parameter, $tA$. The estimate of the observable GRBs is performed as detailed in section \ref{sec:skyvis}.

\section{Definition of Operational Pointing Directions}
\label{sec:op_point}

% Riscrivere cappello:
% - aggiungendo limite dovuto al natural drift solved with alignment of SC FOV. Con commento su figura 9 ora spostata qui.
% - usare tool per settare FD e design di operations

The definition of the pointing directions of each spacecraft, together with their evolution in time, are fundamental elements in the determination of the scientific performance of the constellation.
In fact, pointing strategy design has to find a solution that, while minimizing the number of the required attitude maneuvers and the control effort, allows the coverage of distinct and not obstructed regions of the sky with the overlap of $N>3$ FOVs. The alignment is performed to guarantee the investigation of the celestial sphere on a certain sky area for a minimum amount of time. 

Beside simple pointing strategies, such as zenith and fixed inertial pointing, optimized pointing strategies can be implemented, aiming to determine the pointing directions that maximizes the scientific outcome of the mission.
In this way, the definition of sequential operational pointing directions comes directly from the sky visibility analyses and it is inherently connected to the scientific return of the mission. 

\begin{figure}[tb]
    \centering
    \begin{subfigure}[t]{0.75\textwidth}
    \includegraphics[width=\textwidth]{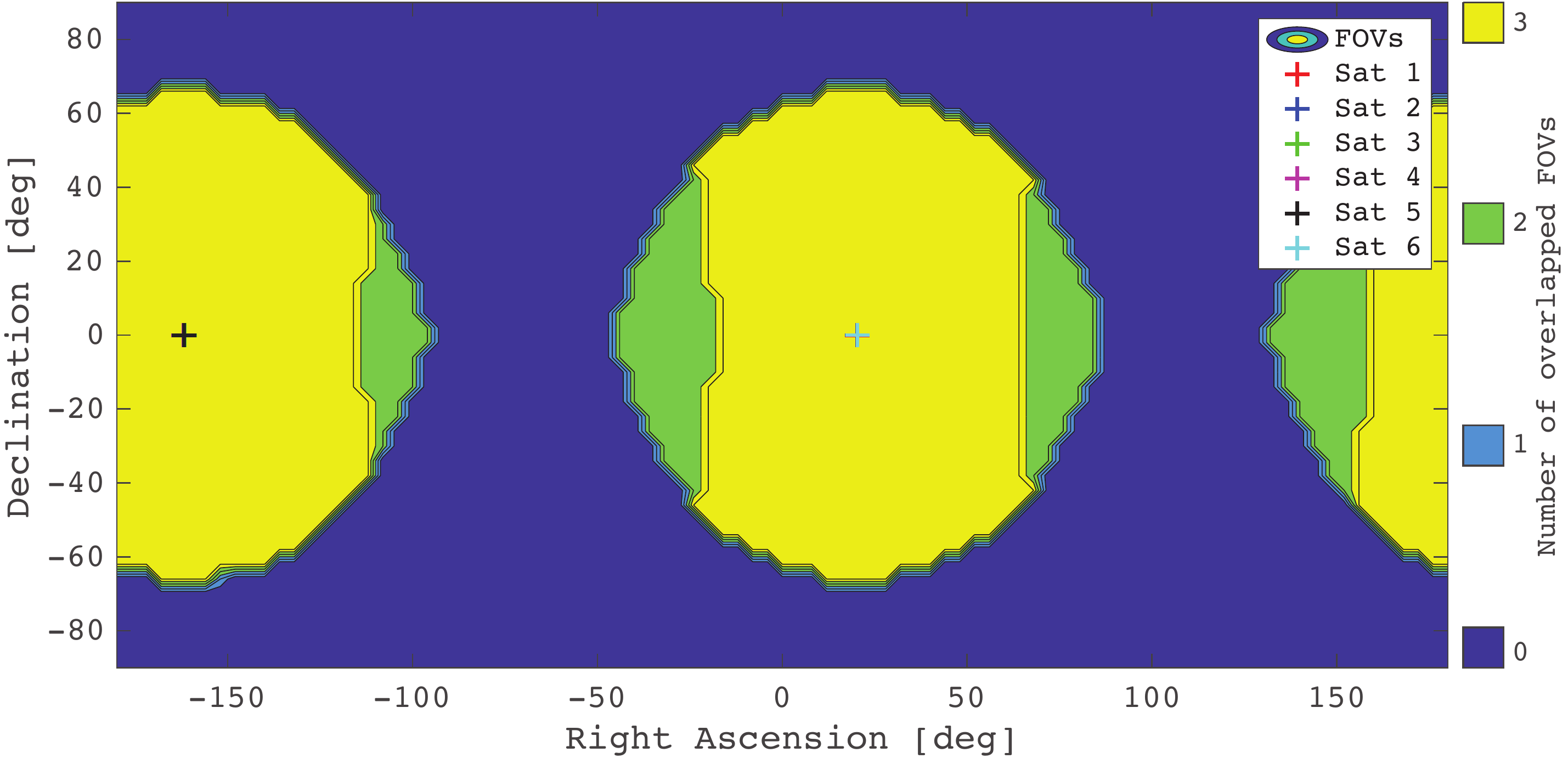}
    \caption{Constellation FOV at $t_0$.}
    \end{subfigure}
    
    \begin{subfigure}[t]{0.75\textwidth}
    \includegraphics[width=\textwidth]{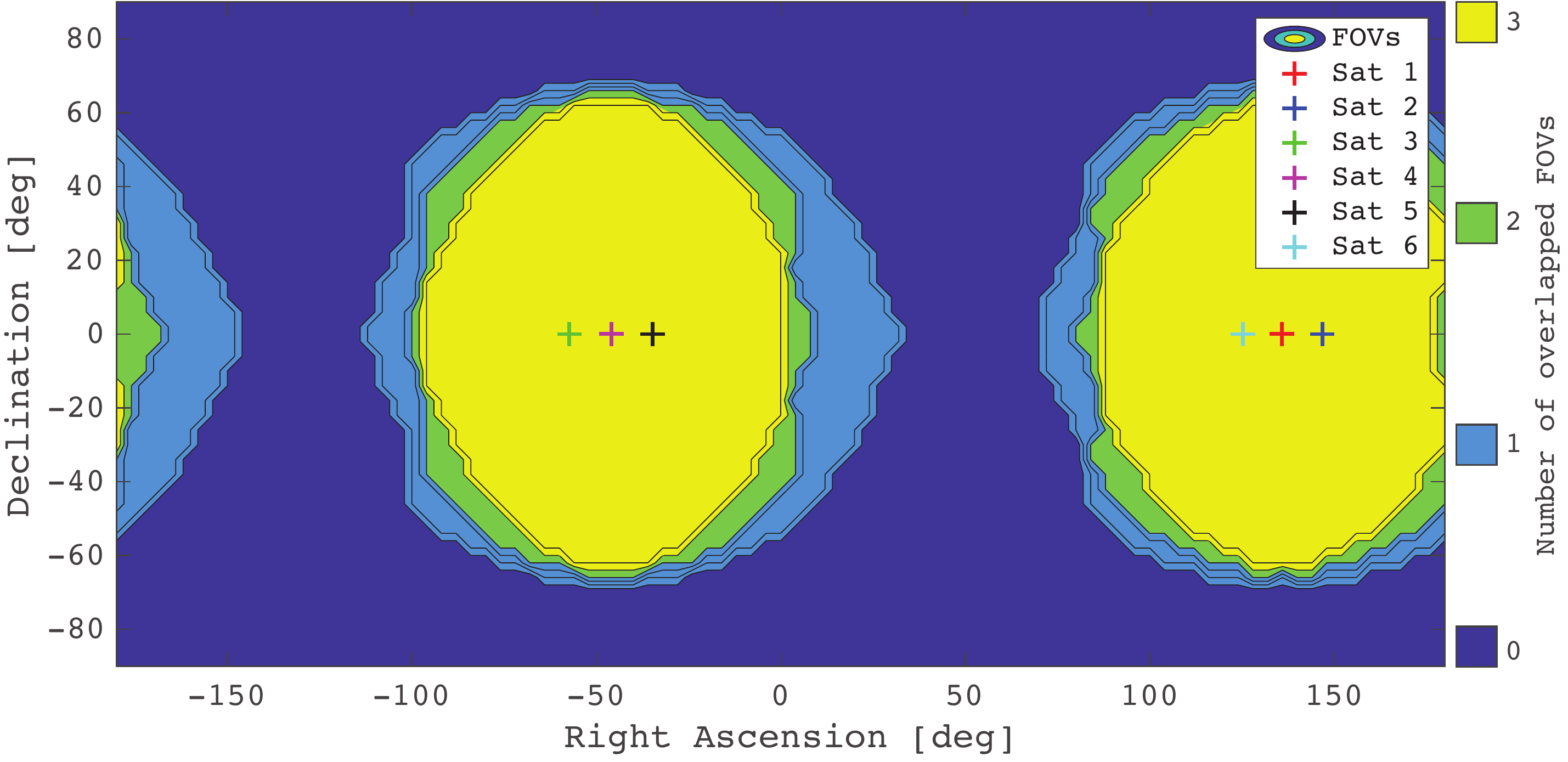}
    \caption{Constellation FOV at $t_0+\SI{7}{d}$.}
    \end{subfigure}
    \caption{Constellation FOV evolution along one week for a LVLH pointing direction.}
    \label{fig:LVLH_fovs_evolution}
\end{figure}

The adoption of time-varying optimized pointing directions is proposed to overcome the limitations given by the lack of orbital control. In fact, the natural relative drift between the elements of the constellation determine a temporal evolution also in the overlap of the FOVs aligned with respect to a LVLH direction, as evident from figure~\ref{fig:LVLH_fovs_evolution}. Therefore, periodic alignments of the different FOVs are required. Ideally, a continuously aligned strategy would be the best option, but this is not feasible due to the limitations inherent with a small space system. In alternative, alignment of the distinct FOVs on inertial directions could provide a different solution to the same issue. However, inertial directions are periodically covered by the Earth surface, leading to a detriment in scientific performances. Moreover, also in this case, periodic alignments on inertially defined regions of the sky are necessary.

Dedicated optimization algorithms have been developed to define the best set of operational pointing directions according to the configurations of the constellation along its dynamical evolution. The frequency of the attitude maneuvers, needed to vary the pointing directions, is not constant and it can be selected by the constellation designer and operators, according to the available attitude management resources and to the required scientific return.
The developed optimization algorithms support the constellation design and the baseline scenario selection. However, their capabilities will be also extensively used during the mission operations as a part of the flight dynamics software to set-up the pointing sequence schedule.

% A heuristic approach has been adopted to develop both optimization algorithms because of the topology of the constraints on the sky and because the speed and the robustness of the method had to be improved. 
Given the topology of the constraints on the sky, a heuristic approach has been adopted for the optimization. Indeed, the constraints on the sky lead to a non mono-convex domain, which means that the domain is a collection of disjoint sets.
The selected algorithm is the MATLAB built-in Particle Swarm, with a swarm size of 20 elements and a maximum number of iteration equal to 100, in order to quickly obtain near-optimal solutions \cite{PSO}. The best sub-optimal solution is then locally-optimized using the MATLAB \textit{fmincon} function, which implements an interior-point algorithm to reach the closest local minimum \cite{HYBRID}.
The numerical effort required to estimate the scientific performances of the constellation with a proper spatial resolution of the sky, using the approach described in section \ref{sec:Sky_Visibility_Tool}, is quite computationally intensive. Thus, it is very time consuming to perform the optimization directly on the actual estimate of the triangulated GRBs. Hence, in both optimization algorithms, only a representative subset of the available instantaneous visibility data has been used. This subset includes a finite number of orbital positions of the constellation within the optimization time period, which correspond to different instantaneous drifted sensors positions with the respective constellation FOV configurations.

\paragraph{LVLH Optimal}
\label{sec:LVLH_optimal}
The working principle of the LVLH optimal strategy is to periodically optimize the direction of the LOS of each satellite with respect to its own rotating LVLH reference frame. 
% As already explained in the previous paragraph, in order to have a fast optimization to be also used during the operations in the flight dynamics software.
As mentioned, it is possible to use a small number of instantaneous visibility data points to estimate with enough precision the effectiveness of a given pointing configuration over a short period (e.g. few days for one week optimization). This feature, which reduces the computational time, is beneficial for the integration of such optimized tool in the flight dynamics software for operation planning.
Indeed, due to the slow behavior of the relative dynamics between satellites, which is in the order of $80-100$ days, and due to the fact that the Earth is fixed in the LVLH frame, the constellation FOV evolution is slow and it is not dependent from the orbital positions of the spacecraft. Figure~\ref{fig:LVLH_fovs_evolution} shows the evolution of the constellation FOV along one week of simulation, from a generic $t_0$ to $t_0+\SI{7}{d}$. It can be noted how the instantaneous observed area of the sky has little differences, and the only significant time evolution of the FOV is contained in the orbital plane of the spacecraft with a fast periodicity of $2\pi$ scan every one orbit.
% Thus, the optimization problem can be solved exploiting only few instantaneous visibility data points spaced by an amount of time that is dependent from the optimization window (e.g. few days for one week optimization).

In order to efficiently exploit this reduction of the optimization computational costs, it is convenient to remove, during the estimation of the triangulable GRBs, the constraints related to the ground-track of the satellites (e.g. large radiations flux regions). Indeed, the dynamics of this constraint is too fast to be correctly identified and to become part of the optimization process. Moreover, the passages of the satellites in forbidden regions is only barely influenced by the pointing direction.

\begin{figure}[tb]
    \centering
    \includegraphics[width=0.75\textwidth]{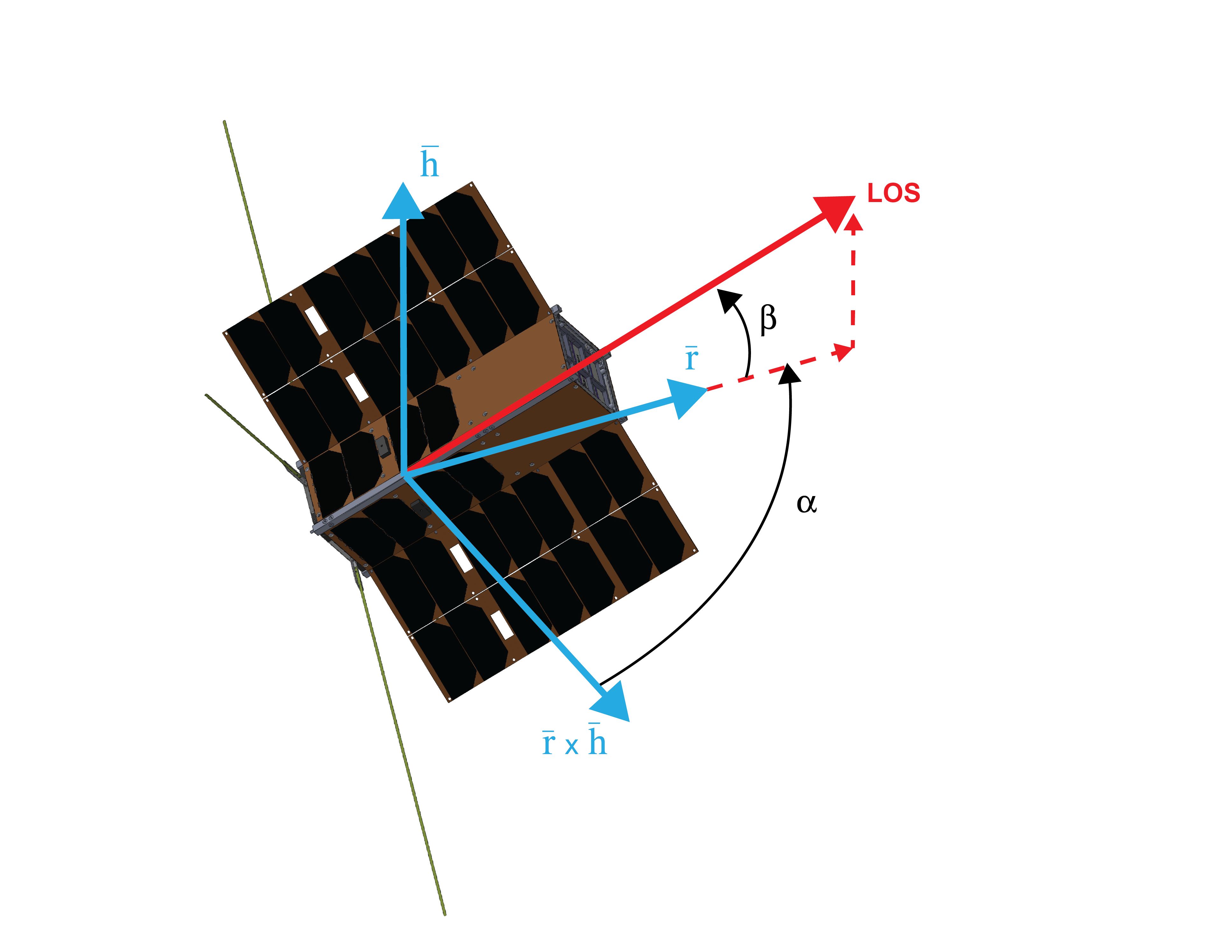}
    \caption{LOS definition in LVLH reference frame.}
    \label{fig:LVLHangle_def}
\end{figure}

With more details on the presented example cases, the algorithm takes into account three points for each optimization window (i.e. the first, the median and the last point of each optimization period) and computes the mean triangulated fraction of the sky. It uses two angles for each satellite, $\alpha$ and $\beta$ in figure~\ref{fig:LVLHangle_def}, as optimization variables to define every LOS in the LVLH frame during the optimization time window. Therefore, for 6 satellites, 12 optimization degrees of freedom are exploited. The optimization algorithm employs 5 to 15 seconds for each optimization period, depending on the satellite configuration, on a 2.6 GHz Intel i7 processor, with no parallel computing technique.

% In order to reduce the complexity of the spacecraft operations only LVLH fixed pointing direction have been considered during each period. 
% Obviously, the shorter is the optimization window the higher is the number of GRBs triangulated, thanks to a frequent optimization of the configuration, but a larger number of attitude maneuvers is required.

It is interesting to underline that this pointing strategy includes also one particular solution in which all the spacecraft are zenith pointing. Nevertheless, this is typically not the optimal one, since a beneficial overlap of the spacecraft fields of view is not possible when the spacecraft have enough physical baseline.

\paragraph{Inertial Optimal}
\label{sec:inertial_optimal}
The working principle of the inertial optimal strategy is to periodically optimize the direction of the LOS of each satellite with respect to the inertial reference frame. Differently from the strategy presented in section \ref{sec:LVLH_optimal}, the Earth is moving with respect to the FOV of each satellite with a period of one orbit (i.e. $\sim \SI{90}{min}$). Therefore, the dynamics of the constraints is not slow enough to allow the optimization on a reduced set of equally spaced data, as in the LVLH optimal case. Anyway, the relative orbital dynamics between satellites is not correlated with the pointing strategy and, thus, shows the same slow dynamics presented in section \ref{sec:LVLH_optimal}. In order to match both the fast and the slow dynamics of the constraints, the optimization of the LOS direction of each satellite is performed using a more refined time grid over one orbit (e.g. 18 points in one orbit, for the presented example cases). 

The logic of the optimization algorithm is similar to the one presented before, with the difference that the optimization variables, $\alpha$ and $\beta$, are defined with respect to the inertial reference frame and the number of points in the optimization window is larger. The optimization algorithm requires 20 to 50 seconds for each optimization period, depending on the satellite configuration, on a 2.6 GHz Intel i7 processor, with no parallel computing technique.
% Also in this case, the slow relative dynamics between the satellites gives the possibility to maintain the optimal inertial pointing direction fixed for a certain period. 
% In the results presented in section \ref{sec:Mission_Analysis_Results}, the observation period with fixed inertial direction is one week.
The advantage of the inertial optimal solution is related to the possibility to easily implement and satisfy pointing direction requirements or constraints on the sky. The pointing requirements or constraints are typically limiting static inertial regions of the celestial sphere, which, in turns, are time varying if projected in the LVLH frame. 
% For instance, the galactic bulge is static in the inertial frame, hence the pointing constraint is simply one direction when pursuing inertial pointing. Instead, if projected on the LVLH, such constraint would be time-varying and consequently more complex to include. 

\subsection{Influence on Astrophysical Data Return}
\label{sec:influence_ap_data_return}

The two different optimized operational pointing strategies have a direct influence on the astrophysical data return in terms of accuracy of the triangulation, coverage of the sky regions and number of triangulated GRBs. The latter is also influenced by the frequency of pointing directions variation, related to the number of necessary attitude maneuvers, which is also discussed in this section.

\begin{figure}[tb]
    \centering
    \begin{subfigure}[t]{0.49\textwidth}
    \centering
    \includegraphics[width=\textwidth]{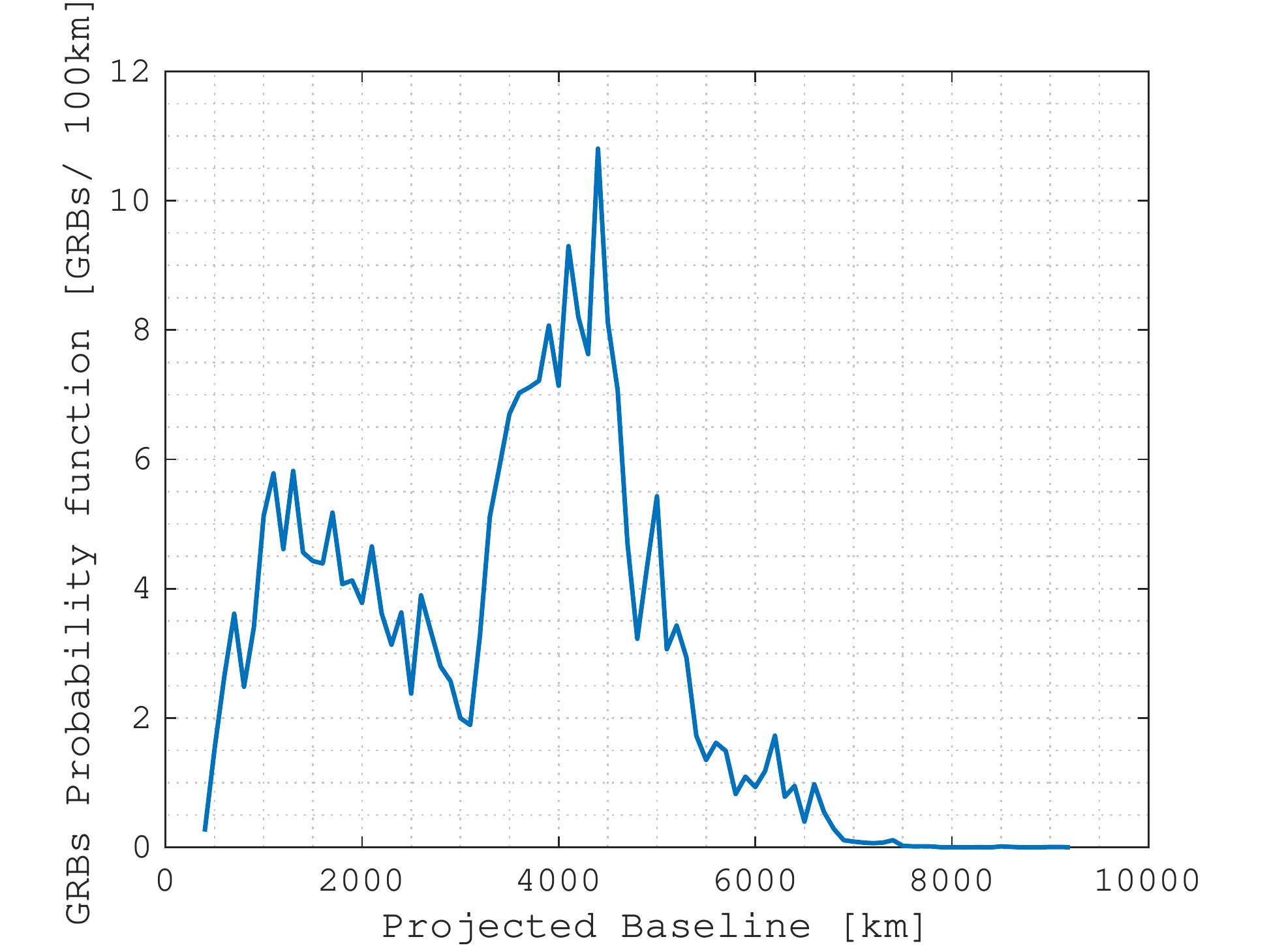}
    \caption{Probability density function.}
    \end{subfigure}
    \begin{subfigure}[t]{0.49\textwidth}
    \centering
    \includegraphics[width=\textwidth]{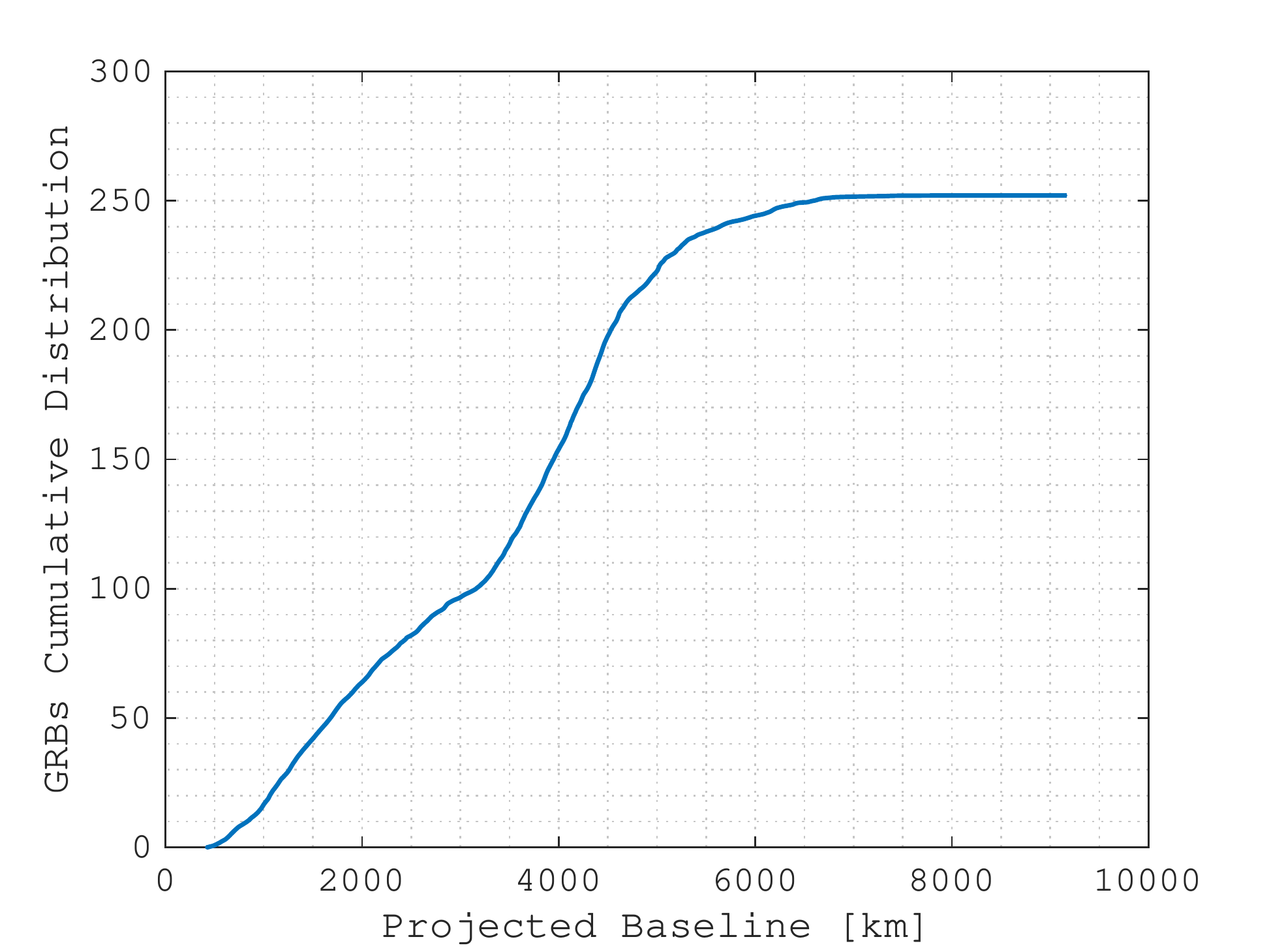}
    \caption{Cumulative density function.}
    \end{subfigure}
    \caption{Projected baseline distribution for LVLH optimal pointing triangulation.}
    \label{fig:baseline_LVLH_opt}
\end{figure}

\begin{figure}[tb]
    \centering
    \begin{subfigure}[t]{0.49\textwidth}
    \centering
    \includegraphics[width=\textwidth]{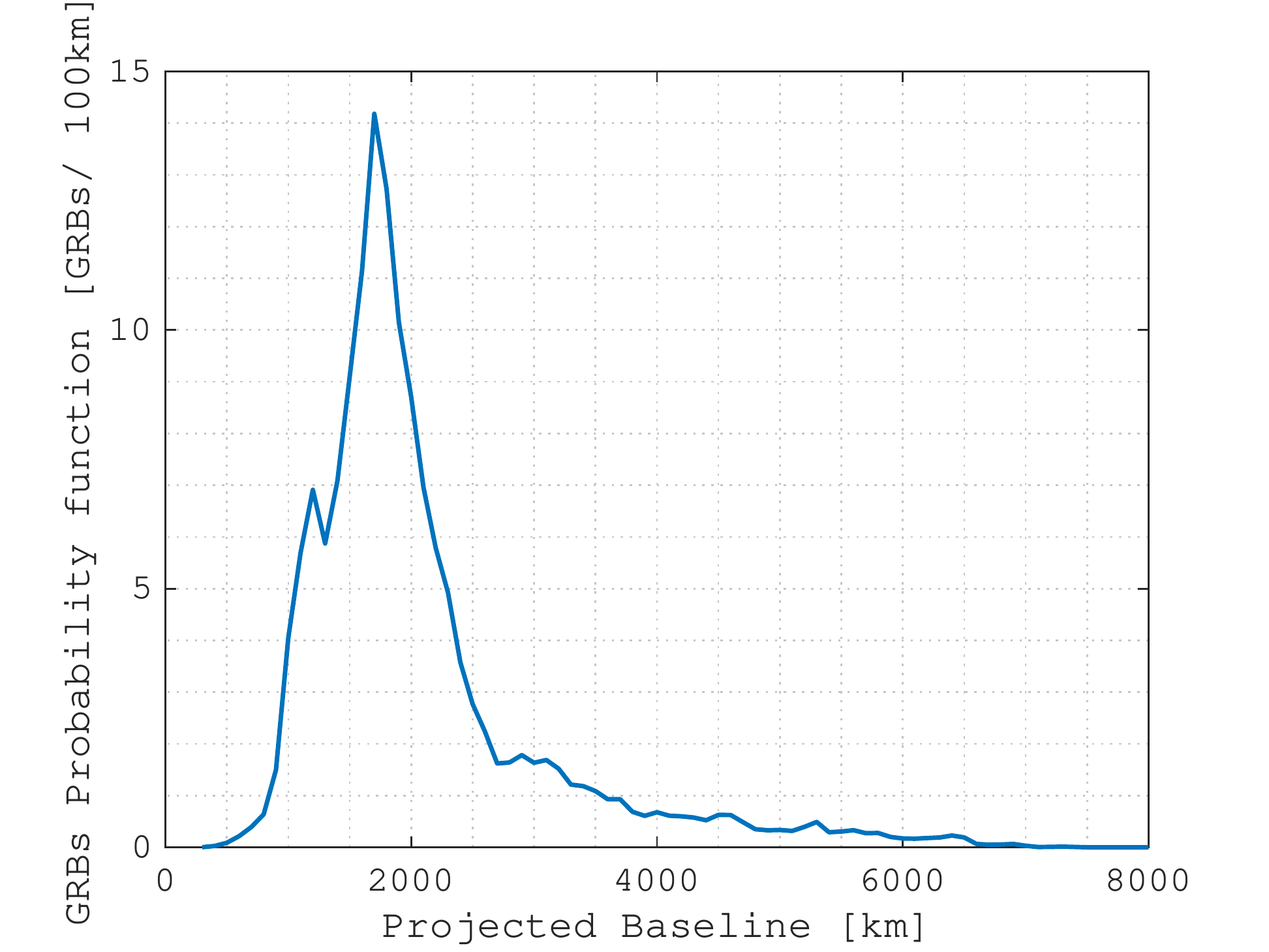}
    \caption{Probability density function.}
    \end{subfigure}
    \begin{subfigure}[t]{0.49\textwidth}
    \centering
    \includegraphics[width=\textwidth]{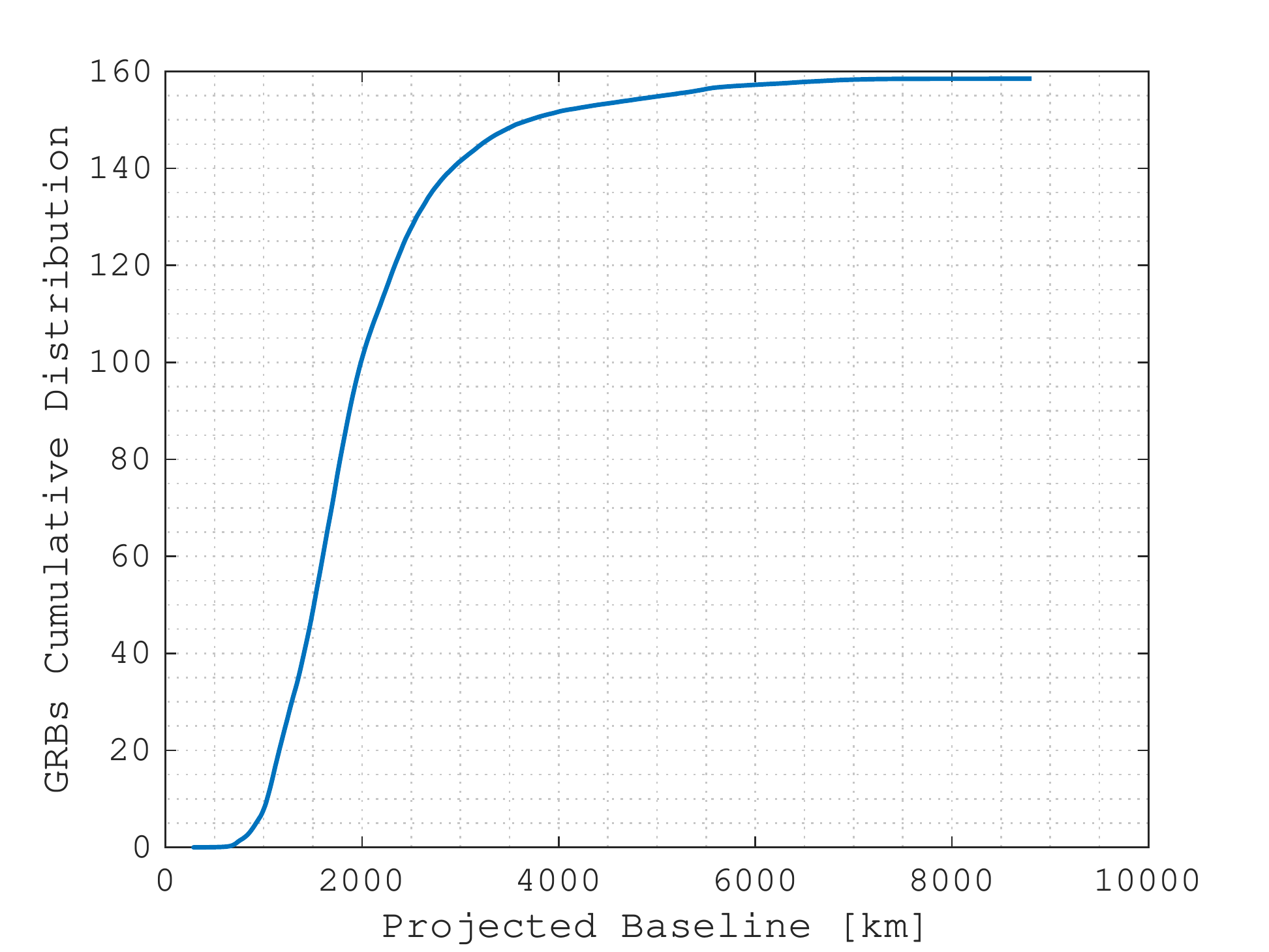}
    \caption{Cumulative density function.}
    \end{subfigure}
    \caption{Projected baseline distribution for inertial optimal pointing triangulation.}
    \label{fig:baseline_INRT_opt}
\end{figure}

Triangulation accuracy is related to the baseline value of the observations. In fact, even if the required minimum baseline is $\SI{1000}{km}$, the larger the projected baseline value during triangulation, the better is the accuracy of the measurements.
Figure~\ref{fig:baseline_LVLH_opt} shows the statistical distribution of the triangulated GRBs as a function of the projected baseline for the LVLH pointing strategy. Since most of the GRBs are observed with a baseline between $\SI{1000}{km}$ and $\SI{5000}{km}$, which is larger than the required one, this pointing strategy leads to a good accuracy in locating the astrophysical events in the celestial sphere.
Figure~\ref{fig:baseline_INRT_opt} shows the statistical distribution of the triangulated GRBs as a function of the projected baseline for the inertial optimal pointing strategy. The typical value for the observation baseline (i.e. $\sim \SI{2000}{km}$) is lower with respect to the LVLH optimal pointing strategy. Therefore, the localization of the GRBs in the sky is less accurate for inertial optimal pointing, with respect to the LVLH one.

% Differences on Sky coverage
\begin{figure}[tb]
    \centering
    \begin{subfigure}[t]{0.49\textwidth}
    \centering
    \includegraphics[width=\textwidth]{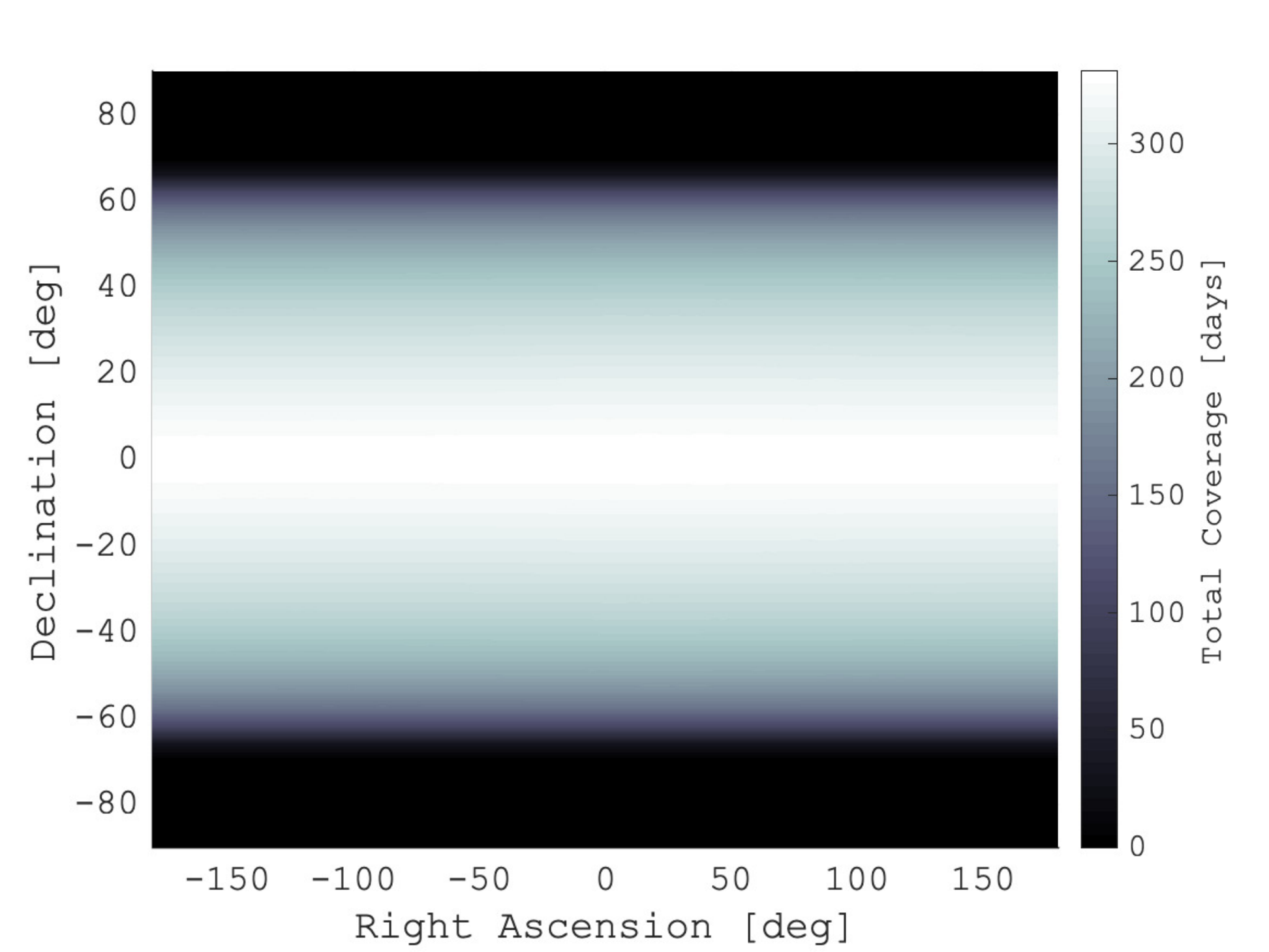}
    \caption{LVLH optimal pointing strategy.}
    \end{subfigure}
    \begin{subfigure}[t]{0.49\textwidth}
    \centering
    \includegraphics[width=\textwidth]{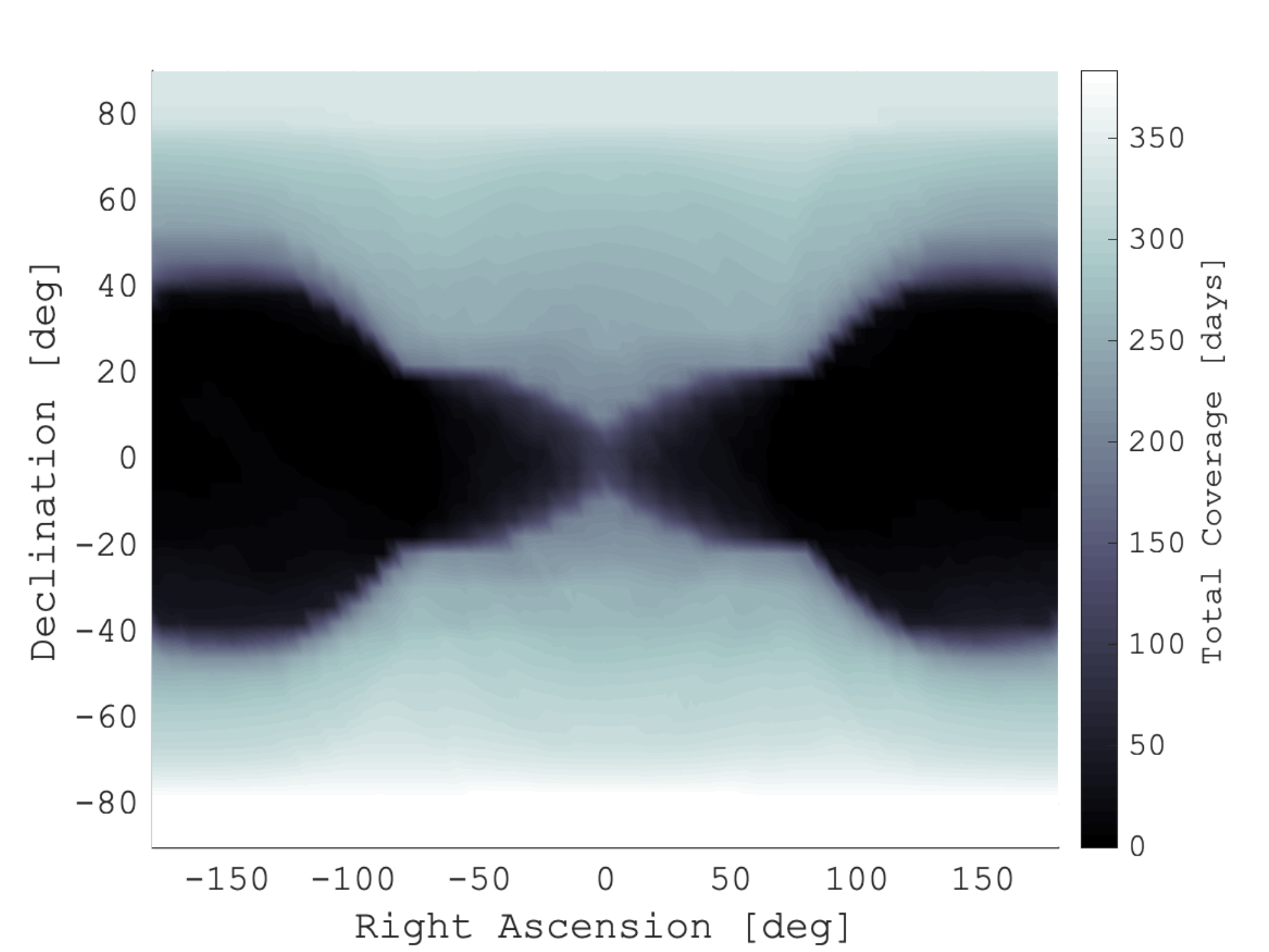}
    \caption{Inertial optimal pointing strategy.}
    \end{subfigure}
    \caption{Sky coverage during the two years mission on equatorial orbits.}
    \label{fig:sky_coverage}
\end{figure}

Figure~\ref{fig:sky_coverage} shows the coverage of the sky for the whole two years mission for both inertial and LVLH-optimal pointing strategies on equatorial orbits. The adoption of different pointing strategies leads to different coverage of the sky: while the inertial pointing strategy promotes the visibility of the polar regions, the LVLH-optimal is more suitable to triangulate GRBs in the equatorial region of the celestial sphere, namely from $\SI{-60}{deg}$ to $\SI{60}{deg}$ of declination. This result highlights how the current design provides a partial analysis of the whole celestial sphere. Hence, to have an investigation over the complete sky catalog, the current constellation design shall be extended, as will be discussed in section~\ref{sec:FC}.

% Number of observed GRBs and maneuvering frequency
\begin{figure}[tb]
    \centering
    \includegraphics[width=0.75\textwidth]{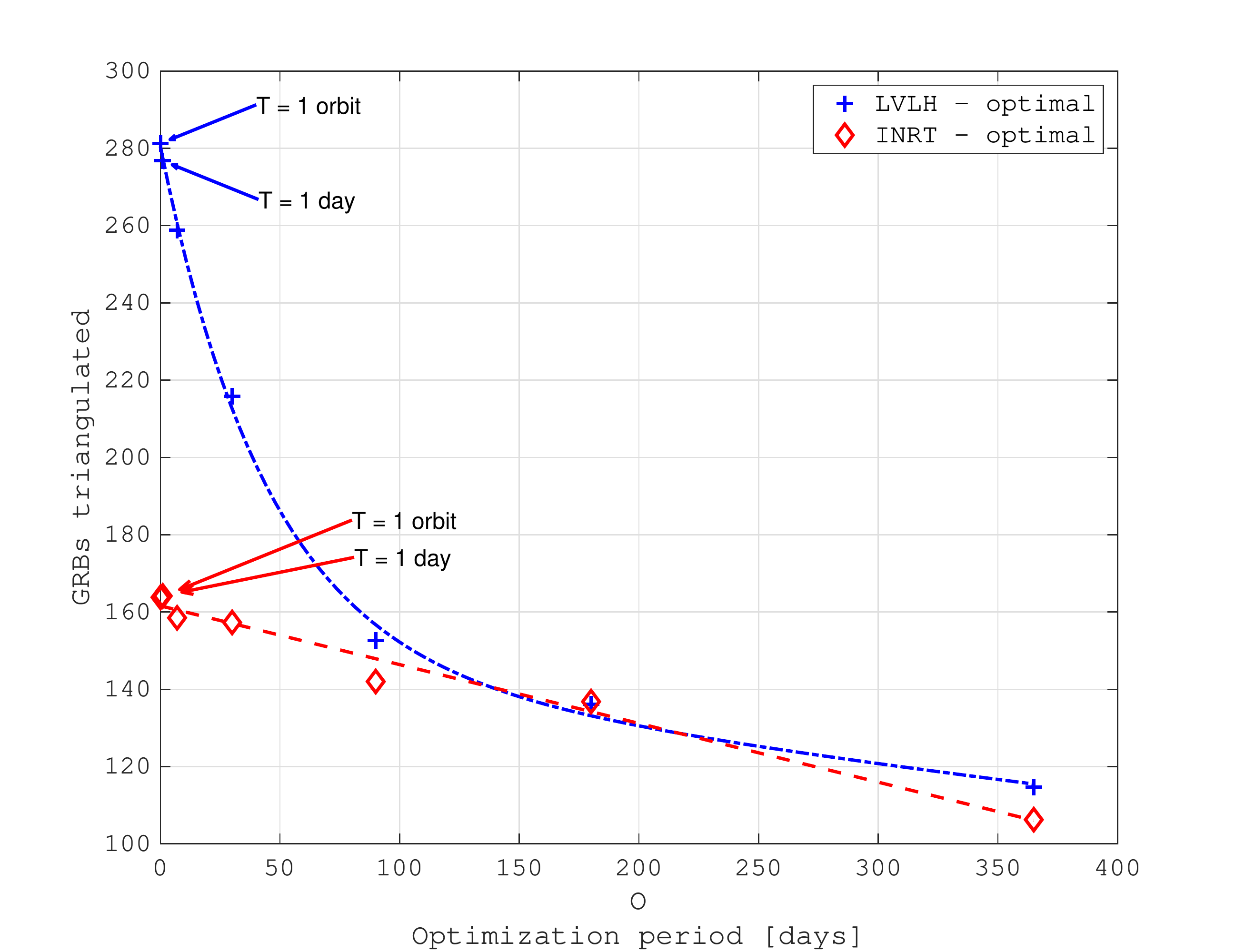}
    \caption{Influence of the pointing update period on the scientific performances.}
    \label{fig:T_opt}
\end{figure}

The selected pointing strategy has a direct impact also on the raw number of potentially observable astrophysical events. In fact, as evident from figure~\ref{fig:T_opt}, the number of triangulated GRBs is directly dependent from the specific pointing of the constellation. However, the definition of the coordinated LOSs directions is not the only pointing design parameter; the number of alignment attitude maneuvers plays a non-negligible role in determining the scientific output of the mission. In general, since the optimized pointing strategies are evaluated over a certain time window, it is easily arguable how more frequent attitude maneuvers, to periodically align the different FOVs, lead to a finer optimization grid and, thus, to better scientific performances. On the contrary, the consequent load on the system platform may be not sustainable within the limitations of a small space system. In these regards, figure~\ref{fig:T_opt} shows the influence of the maneuvering frequency on the scientific results for both LVLH and inertial pointing. As expected, for both strategies, the higher the optimization frequency (i.e. lower pointing update period), the better the scientific return. If the pointing update period is longer than the period of the relative motion (i.e. $80-\SI{100}{d}$), the two strategies show similar results. In this case, the rare pointing optimization maneuvers do not allow to overcome the limitations due to the natural drift of the constellation. The benefit in selecting specific optimal directions is marginal.
Conversely, if the pointing update period is reduced, the LVLH optimal pointing strategy is able to ensure a higher gain in the number of triangulated GRBs. This strategy beneficially exploits a frequent optimization of the alignment between the constellation elements. In this way, the best FOV overlap is available at any time the satellites are satisfying the position requirements.
This is not the case for the inertial optimal strategy since, even with ideal continuous pointing optimization, some areas of the sky are periodically behind the Earth sphere for certain elements of the constellation. Hence, there is an upper limit in the possible scientific gain. 

\section{HERMES Constellation Scenario}
\label{sec:Mission_Analysis_Results}
% The proposed method and analytical instruments can be generalized to any kind of distributed space mission with $N$ satellites observing the sky, whose performance are dependent from the relative dynamics between the space elements and from the coverage of their fields of view. Nevertheless, in this particular context, the presented analysis is focused on determining the most performing mission scenario for HERMES constellation, in terms of nominal orbit, injection and pointing strategy.
The proposed method has been employed to design and analyze the effectiveness of HERMES constellation.
In particular, to consolidate the results with respect to unavoidable injection and release uncertainties, dedicated Monte Carlo statistical simulation campaigns are used.

% Preliminary analyses are necessary to determine the nominal mission scenario. Their details are out of the scope of this paper, since they are strongly characterized by the current mission details, such as the launch date and window, or the technical specifications of the deployer. However, the selected nominal mission scenario is described in the following section. Once the nominal parameters are set, a comprehensive statistical analysis can be carried out to assess the robustness of the proposed mission scenario and the confidence on expected scientific performances, as a function of the unavoidable injection uncertainties.

\subsection{Mission Scenario Definition}
\label{sec:misscen}
% The nominal mission scenario is characterized by an operational orbit, a pointing strategy and an injection strategy. 
The possible operational orbits, according to the scientific requirements discussed in section \ref{sec:scireq}, are nearly-equatorial orbits or Sun Synchronous orbits with altitude between $\SI{450}{km}$ and $\SI{600}{km}$. 
% A positive correlation between the altitude of the orbit and the scientific performances exists, resulting in slightly better results at higher altitudes. 
Due to orbit injection errors, a nominal orbit with altitude of $\SI{500}{km}$, or lower, does not guarantee the $\SI{2}{y}$ mission duration requirement in all the possible cases, with enough confidence. Hence, a nominal orbit altitude of $\SI{550}{km}$ is selected, in combination of a nominal eccentricity of $0$ (i.e. circular orbit). 
% In this way, the altitude of the satellites is within the acceptable limits in $99.73\%$ of the possible cases: the mission duration and the radiation environment requirements are satisfied.
It is worth noting that the presence of large radiation flux regions at the poles of the Earth, as specified in figure~\ref{fig:earth_flux}, reduces the available observation time for polar and nearly-polar orbits. Indeed, the available estimates of the scientific performances on SSO are tremendously reduced with respect to those on low inclination orbits. Hence, the nominal scenario is set on equatorial orbits, with $i = \ang{0}$.

%However, the output of the investigations about the scientific results are applicable for the entire class of low inclination orbits (i.e. $\ang{0}\le i \le \ang{30}$), as specified by the scientific requirements. 
% The nominal operational orbit is a low nearly-equatorial circular orbit with $h=\SI{550}{km}$.

%% NOTA IL PARAGRAFO PRECEDENTE SI PUO' RIDURRE MOLTO

\begin{figure}[tb]
    \centering
    \includegraphics[width=0.75\textwidth]{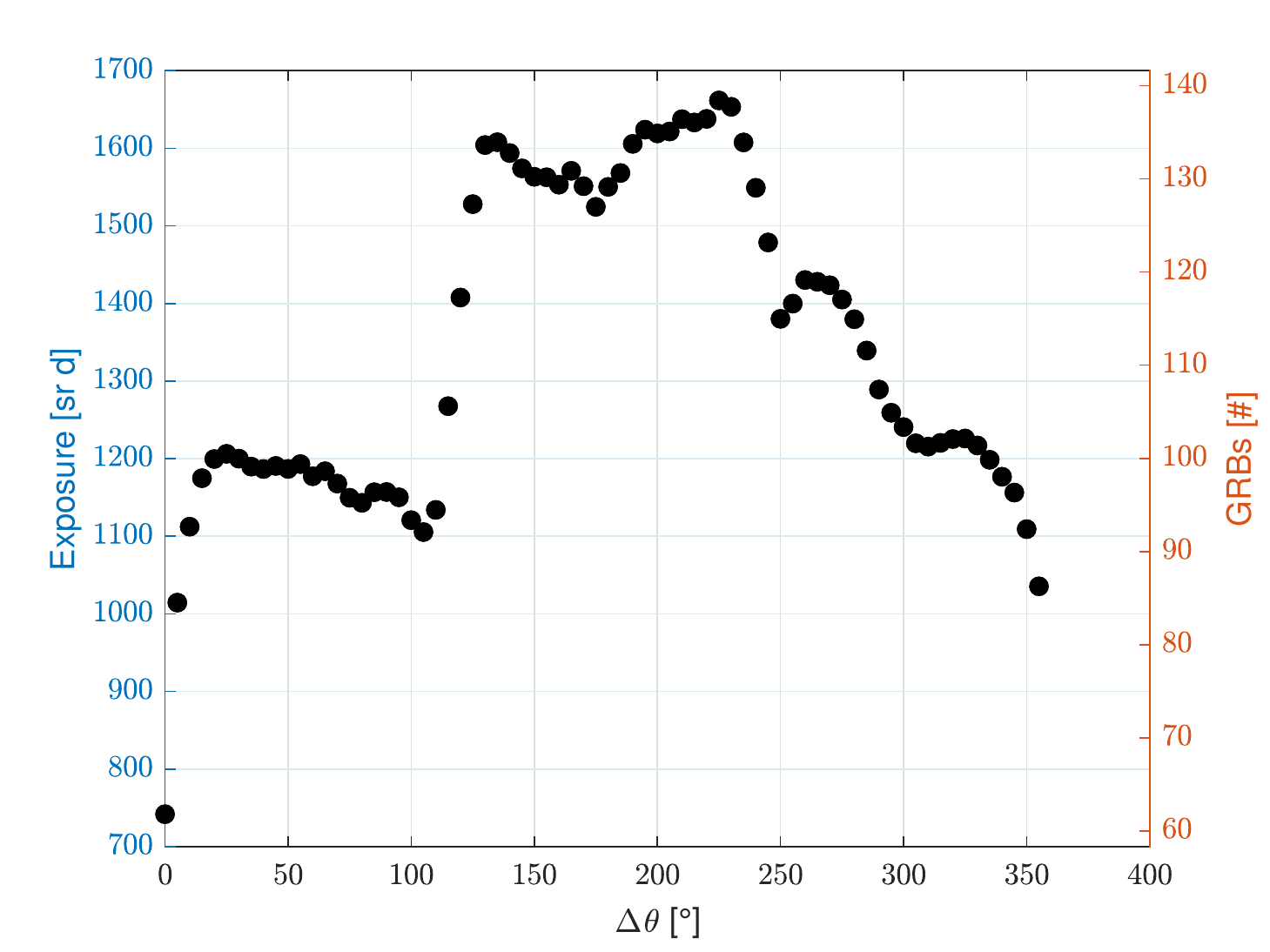}
    \caption{Estimate of the scientific performances as a function of the phasing angle in the true anomaly separation for two triplets injected at $h=\SI{550}{km}$. The figure reports both the exposure, $tA$, and potentially observable GRBs.}
    \label{fig:prelim_phasing}
\end{figure}

For what concerns the injection strategy, the single injection of multiple spacecraft is considered to be favorable both for the greater performances in terms of launcher availability, and for the lower sensitivity with respect to release condition errors, as extensively discussed in section \ref{sec:const_design}. Hence, this is assumed as the nominal injection condition for the considered mission scenario. The remaining free variable is the phasing angle between the release of the two triplets in the constellation. 
% In fact, as discussed for the case of the single injection of multiple spacecraft in section \ref{sec:singleinj}, three satellites are injected in a short time frame in a point of the orbit.
The nominal angular separation between the two injections of the triplets has been selected with a preliminary analysis on the possible phasing angles between the two release events. In particular, a phasing angle in the order of $\ang{140}$-$\ang{220}$ results in better scientific performances. This is because, the central elements of the triplets (i.e. spacecraft injected along the h-bar direction) remain always on opposite side of the orbit, guaranteeing the possibility to form an active triplet as soon as a third satellite is available with enough relative distance. The triangulable GRBs as a function of the phasing angle between two triplet injections is reported in figure~\ref{fig:prelim_phasing}. 
%For the sake of simplicity and comparison, the simulations were performed using zenith pointing.

The nominal spring value for the release $\Delta v$ is selected according to the availability of typical nano-satellite deployers. A nominal release $\Delta v=\SI{1.25}{m/s}$ has been selected. In this case, the relative motion is dominated by the spring authority over the natural gravitational perturbations. 
% As a result, the actual location of the spacecraft injection is less relevant for what concerns the expected scientific outcome. 

The selection of the nominal optimal pointing strategy is dependent from the comparison in section~\ref{sec:influence_ap_data_return} and from the robustness analysis discussed in the next section. An example pointing update period of one week is used, being a good compromise between the complexity of the spacecraft operations and the scientific return. 

\subsection{Robustness Assessment with Statistical Analysis}
\label{sec:robustness}

To assess the confidence on the estimate of the scientific performances and the robustness of the proposed mission scenario with respect to unavoidable injection uncertainties, a comprehensive statistical analysis is set up. In particular, a Monte Carlo simulation campaign is used. 

The pool of uncertain variables adopted for the robustness analysis is reported in table~\ref{tab:uncertain_MC}. The selected set includes launcher's and deployer's injection errors. 

\begin{table}[tb]
\caption{Nominal variables, ranges and random errors.}
\centering
\begin{tabular}{lccccc}
\toprule
 & Nominal value&\multicolumn{2}{c}{Design range}&
 \multicolumn{2}{c}{Random error}\\
 & & Range & Distr. & 3$\sigma$ & Distr.\\
 \midrule
 \textbf{Launcher}&&&&\\
Launch Window [y]& 0 &2&Uniform&-&-\\
Phasing [h]& 0 & 48 &Uniform&-&-\\
SMA [km]&6928&-&-&30&Normal\\
e [-]&0&-&-&0.0024&Half-normal\\
i [deg]&0&-&-&0.30&Normal\\
$\omega$ [deg]&0&-&-&1.2&Normal\\
$\Omega$ [deg]&0&-&-&0.4&Normal\\
$\theta$ [deg]&0 and 220&-&-&1.2&Normal\\
 \textbf{Release}&&&&\\
 Spring $\Delta v$ [$\frac{m}{s}$]&1.25&0.25&Uniform&10$\%$ of $\Delta v_{nom}$ &Normal\\
 Injection angle [deg]&0&-&-&4&Normal\\
\bottomrule
\end{tabular}
\label{tab:uncertain_MC}
\end{table}

As far as the launcher is considered, both temporal and orbital uncertainties are taken into account. In particular, the time variability is represented by a 2 years wide launch window (i.e. the launch date is within 2 years from 1st of April 2020) and a 2 days wide phasing window (i.e. the second triplet is released in a different point of the orbit within 48 hours from the first triplet). It is worth noting that the phasing window can be extended to a longer time-frame without significant changes in the results. %In fact, the results are mainly dependent from the angular shift of the two triplets along the orbit, which is a design parameter and can be obtained despite the actual phasing time between the two triplets. 
The orbital uncertainty of the launch is represented with errors in all the 6 Keplerian parameters of the nominal orbit. 
The reference values for launcher injection uncertainties come from the combination of the worst cases among the European launchers (e.g. Vega, Ariane 5, Soyuz), with a margin of $100\%$. 

In addition, the analysis considers the uncertainties for the release from the deployer. In fact, after each triplet is injected with uncertainties into the orbit, the spring release of each satellite from the deployer is taken into account. The three satellites of each triplet are released in three different LVLH direction to achieve the desired relative motion.  The spring release is affected by errors in both the magnitude and direction of the spring $\Delta v$, as listed in table~\ref{tab:uncertain_MC}. The injection angle error is taken from the worst cases among the European launchers in terms of upper stage attitude control accuracy, margined by $100\%$. For what concerns the uncertainty on the impulse imposed by the release spring, uniform distribution around the design nominal value of $\Delta v_{nom}=\SI{1.25}{m/s}$ is used. For each run of the analysis a unique uniformly dispersed $\Delta v_{run}$ value is selected within the spring $\Delta v$ range (i.e. spring value for the single run $\Delta v_{run}=\Delta v_{nom}\pm\text{Uniform}$). This is done to prove the robustness also with respect to different spring $\Delta v$s of commercially available deployers.
Then, each satellite is released with the actual spring value, $\Delta v_{sat}$, which is dispersed normally around $\Delta v_{run}$ (i.e. $\Delta v_{sat}=\Delta v_{run}\pm\text{Normal}$).

The number of runs of the statistical analysis is selected to represent the global response of the system to the assumed uncertainties, according to Hanson \cite{MCNasa}. 
The goal of the analysis is to estimate the value of the expected triangulable GRBs by the constellation of nano-satellites. Therefore, in terms of statistical analysis, the goal is to estimate the mean value of the expected GRBs and to bound the scientific output within a box at $\sim99.73\%$ of probability to achieve the scientific result. 
Preparatory analyses showed that with $\sim 100$ samples the standard error of the estimated mean was below $2$ GRBs. In this way, the expected mean number of GRBs can be estimated with a precision range of $5$. 
Pool of variables is created exploiting a Mersenne Twister random number generator with seed continuously shuffled according to code execution time on different servers.

\begin{figure}[tb]
    \centering
    \includegraphics[width=0.75\textwidth]{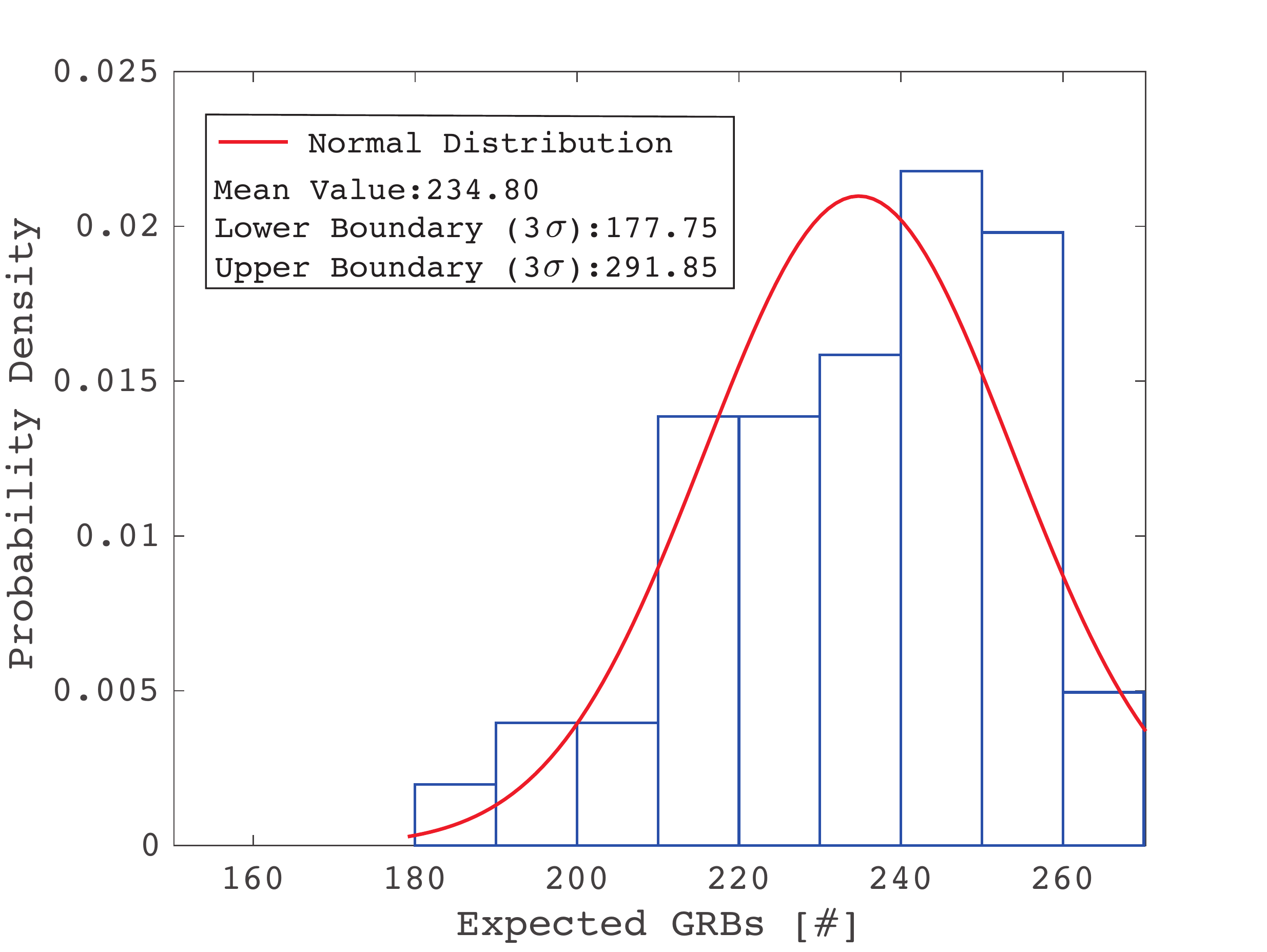}
    \caption{Statistical analysis results for the optimized LVLH pointing. Number of expected triangulable GRBs throughout the mission.}
    \label{fig:mc_lvlh_optimal}
\end{figure}

\paragraph{LVLH Optimal Results}
The results for the statistical analysis employing the LVLH optimal pointing, described in section \ref{sec:LVLH_optimal}, are reported in figure~\ref{fig:mc_lvlh_optimal}.
At $\SI{550}{km}$ nominal altitude, the mean value of expected GRBs is $234.80$, the standard error of the mean is $1.94$ and the standard error of the standard deviation is $1.39$, as reported in table~\ref{tab:robustness_optimized}. Hence, the mean value of expected GRBs is estimated to be between $228.98$ and $240.62$ at $\SI{99.73}{\percent}$.

\begin{table}[tb]
\caption{Robustness analysis results for the LVLH optimized pointing strategy.}
\centering
\begin{tabular}{lcccc}
\toprule
h [km]&$\mu$ [GRB] & 3$\sigma$ [GRB] & $SE_{\mu}$ [GRB]& $SE_{\sigma}$ [GRB]\\
 \midrule
550&234.80&57.03&1.94&1.39\\
\bottomrule
\end{tabular}
\label{tab:robustness_optimized}
\end{table}

\paragraph{Inertial Optimal Results}

\begin{figure}[tb]
    \centering
    \includegraphics[width=0.75\textwidth]{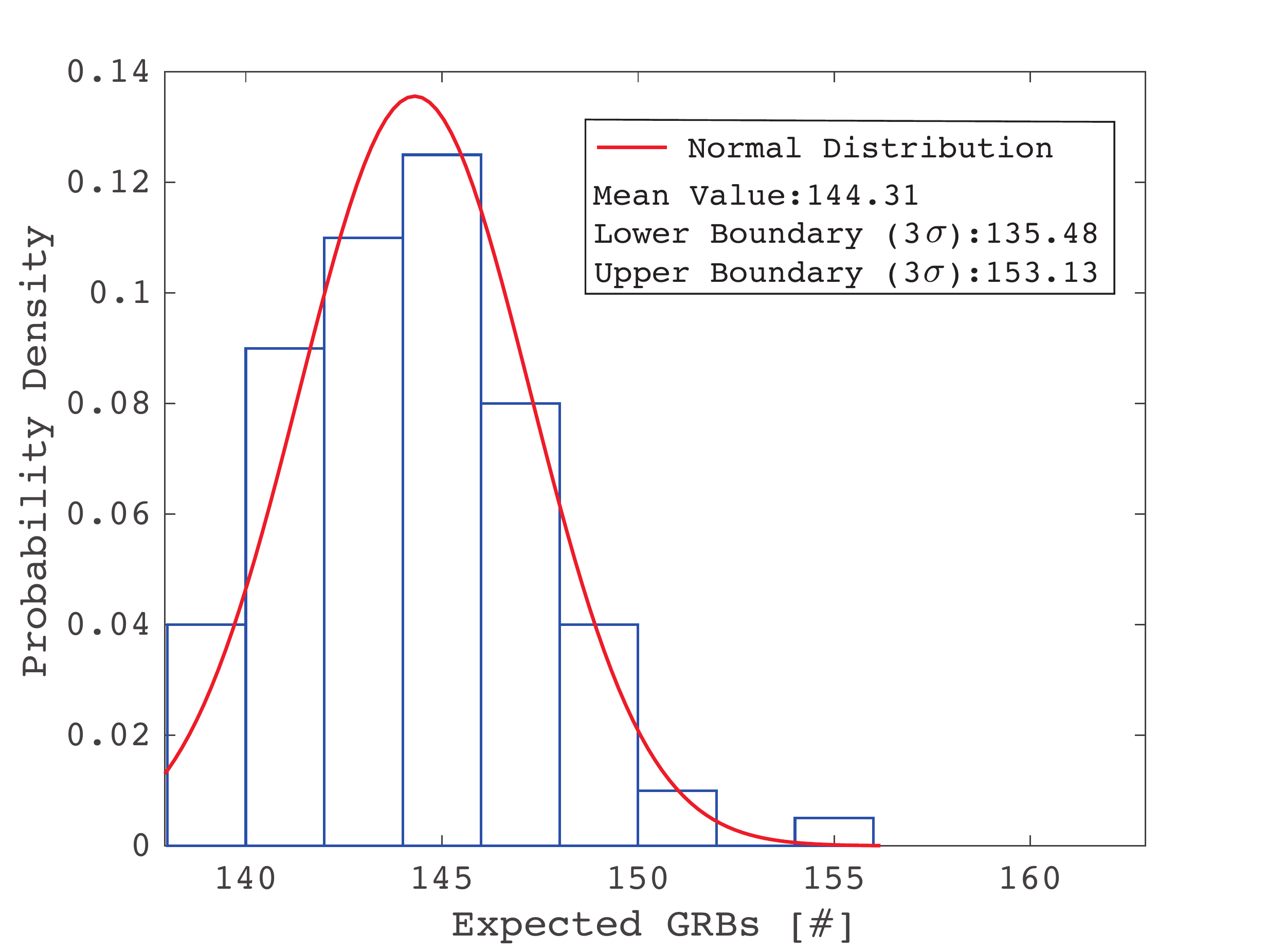}
    \caption{Statistical analysis results for the optimized inertial pointing. Number of expected triangulable GRBs throughout the mission.}
    \label{fig:mc_inertial_optimal}
\end{figure}

The results for the statistical analysis employing the inertial optimal pointing, described in section \ref{sec:inertial_optimal}, are reported in figure~\ref{fig:mc_inertial_optimal}.

At $\SI{550}{km}$ nominal altitude, the mean value of expected GRBs is $144.31$, the standard error of the mean is $0.30$ and the standard error of the standard deviation is $0.22$. Hence, the mean value is estimated to be between $143.41$ and $145.21$ at $\SI{99.73}{\percent}$, as reported in table~\ref{tab:robustness_optimized_inertial}.

\begin{table}[tb]
\caption{Robustness analysis results for the inertial optimal pointing strategy.}
\centering
\begin{tabular}{lcccc}
\toprule
h [km]&$\mu$ [GRB] & 3$\sigma$ [GRB] & $SE_{\mu}$ [GRB]& $SE_{\sigma}$ [GRB]\\
 \midrule
550 & 144.31 & 8.83 & 0.30 & 0.22\\
\bottomrule
\end{tabular}
\label{tab:robustness_optimized_inertial}
\end{table}

\medskip

The statistical analysis presented in this section has been used to prove that the estimate of triangulable GRBs is basically uncorrelated to the uncertain variables taken into account in the mission analysis. This feature is beneficial in terms of mission design, given that the statistical analysis was performed against typical launch and injection uncertainties. Thus, the presented statistical analysis does not explore different mission concepts (e.g. correlation with design variable), rather it guarantees the achievement of nominal results.

The LVLH optimal pointing strategy is confirmed to provide better results than the inertial optimal one, in terms of both number of triangulated GRBs and typical observation baseline. One reason for such result is the FOV occultation due to Earth. In the inertial pointing, Earth enters the FOV of the satellite instruments forming a triplet, alternately. This leads to preferred pointing regions close to the poles, which in turns are restricted by the achievable projected baseline. This degrades the performance in terms of available observation time and observable regions. The lower standard deviation of the inertial distribution may be due to the modest sensitivity to relative motion, differently from the LVLH pointing strategy, which has a temporal drift in the FOV overlap, as discussed in section~\ref{sec:op_point}.
For the above-mentioned reasons, the nominal pointing strategy for the HERMES mission scenario described in section \ref{sec:misscen} is the LVLH optimal.

\medskip
%\subsection{Robustness of the Selected Mission Scenario}
%The variability of the estimates of the available scientific performances is low and within acceptable bounds.  
% Nevertheless, due to launch constraints and uncertainties on the survivability of the satellites, it is important to assess the robustness of the mission to a non-nominal constellation configuration. 

\begin{figure}[tb]
    \centering
    \includegraphics[width=0.5\textwidth]{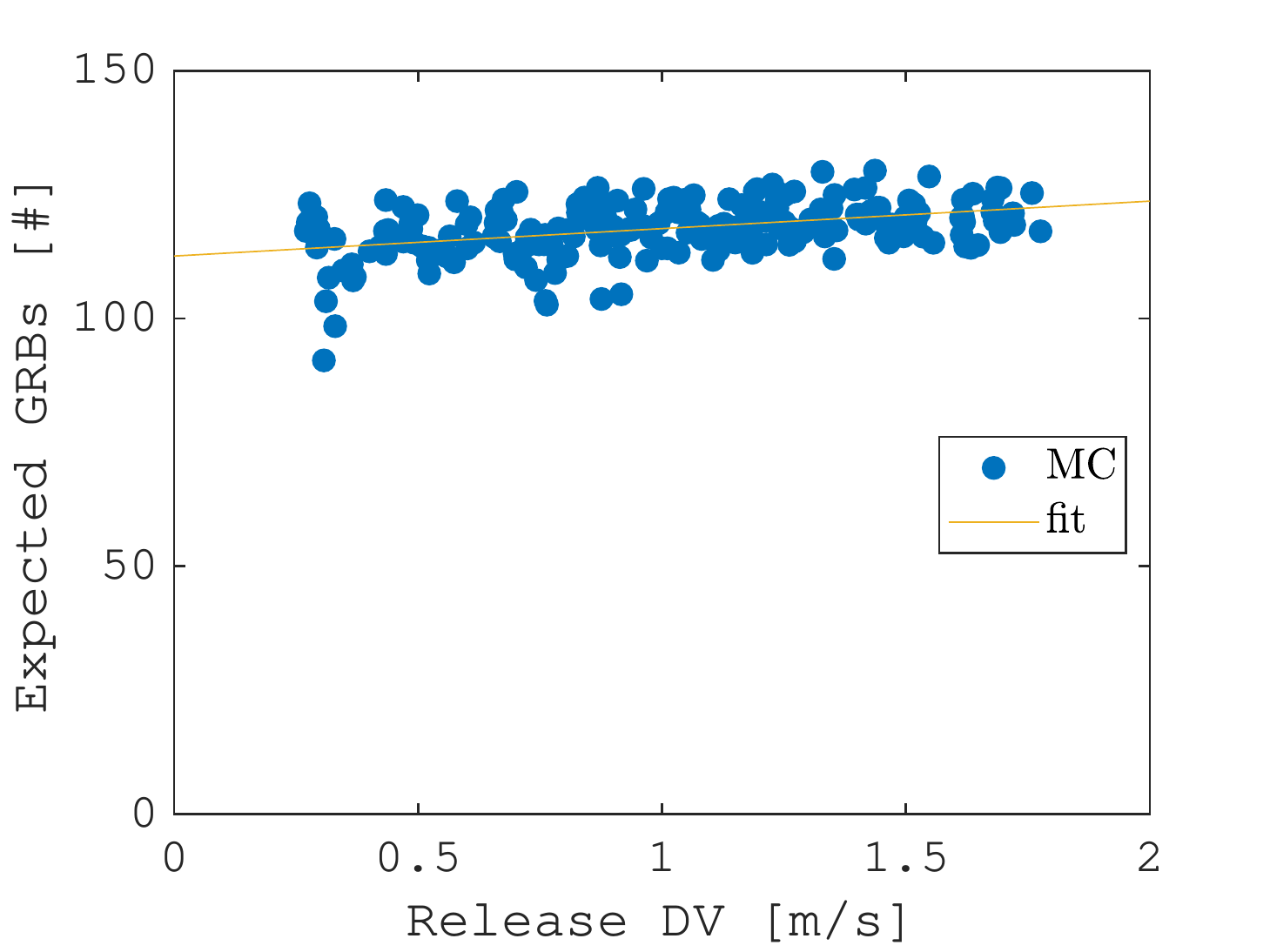}
    \caption{Correlation between potentially observable GRBs and release spring $\Delta v$.}
    \label{fig:correlation_MC}
\end{figure}

As anticipated in the description of the mission scenario, the correlation of deployer spring $\Delta v$ with the scientific outcome has been assessed with a Monte Carlo campaign on a set of 250 samples. % This set of samples was able to highlight correlations between potentially observable GRBs and the mission scenario variables, which helped in targeting the mission analysis to the most feasible and promising scenario as discussed before. 
In particular, a weak correlations is outlined in figure~\ref{fig:correlation_MC}. The Pearson's linear correlation coefficient is 0.43.
In addition, it is here remarked that, as the $\Delta v$ decreases, the dispersion of the expected GRBs increases, and the absolute number decreases. This may be explained by the fact that, for low spring impulses, the relative motion is governed by the differential geopotential acceleration experienced at injection into orbit, hence it is highly sensitive to initial conditions. On the other hand, the relative motion is governed by the imposed impulse whenever higher $\Delta v$ is delivered by the spring.

% As already said before, a weak correlation is found between the orbit altitude and the number of expected GRBs. In particular, the Pearson's linear correlation coefficient is 0.32. An orbital altitude of $550\ km$ delivers higher results in terms of expected GRBs with respect to lower orbit altitudes.

\medskip
The robustness of the mission design is sought also with respect to non-nominal configuration scenarios and eventual satellite failures. 
One potential non-nominal configuration is the presence of a single triplet. Results show that acceptable scientific outcome can be guaranteed even with just 3 satellites, scaling almost linearly the number of expected GRBs regardless of the operational pointing strategy (i.e. $\SI{\sim50}{\percent}$ scientific performances reduction with respect to a constellation of 6 elements). The analysis is useful to evaluate a hypothetical scenario in which the two triplets are launched with a significant in between delay. Indeed, in such configuration, the 6 satellites constellation would exist far less than two years. For instance, assuming the two triplets are launched one year apart, the constellation will function as a single triplet for one year, a 6 satellites constellation for another year and, again, as a single triplet for the last year. 
A further possible non-nominal configuration is a 5 satellites constellation. This is representative of an early failure in a single space element (e.g. one satellite is lost during injection or commissioning phase). The results demonstrate that losing the possibility to form two distinct triplets reduces the performances to almost those available with a single triplet. In fact, at any instant of time, the constellation field of view is analogous to the one available with just three elements. 
However, 5 satellites have better performances than a single triplet.
This is because 5 satellites reduce the time in which triangulation is not possible due to position constraint violation of a single element in the constellation. 
Thus, a satisfactory robustness towards a single satellite failure is present. Indeed, the reduction in the estimate of the scientific performances is in the order of $\SI{\sim40}{\percent}$, with respect to a constellation of 6 elements, for both operational pointing strategies. 

\subsection{Extension to a Full Constellation}
\label{sec:FC}

The present mission design is dedicated to the improvement of the scientific performances within the HERMES pathfinder mission requirements. However, the constellation configuration can be extended to achieve better or more astrophysical data. In fact, the major shortcoming of the current design is the limitation in having all the spacecraft on a single orbital plane. This is motivated by the launch opportunity availability, but does not allow a complete three-dimensional triangulation. In fact, the hemispherical uncertainty, with respect to astrophysical events coming from above or below the orbital plane, cannot be solved by the proposed configuration. Moreover, the selection of the LVLH optimal pointing strategy leads to an incomplete catalog of observable astrophysical events, restricted to those coming from the equatorial section of the celestial sphere, as outlined in section~\ref{sec:influence_ap_data_return}.

A complete sky coverage can be achieved by dividing the two triplets, in a way that one is inertial pointing and, the other, LVLH pointing. Nevertheless, a lower number of triangulated GRBs is expected. It shall be noted that an increase in the frequency of pointing updates, even with a mix of inertial and LVLH directions, does not solve the intrinsic limitations of the present configuration. Additionally, it may over-stress the attitude control subsystem, potentially leading to premature failures of the small space system. Then, the simple enlargement of the constellation in its current configuration is not the best viable solution. 

The proposed extension to a full constellation of small satellites for localization and triangulation of astrophysical events is to have $N_{SC}>6$ spacecraft, subdivided in different orbital planes. The greater part of the constellation shall be located at low inclination (i.e. $ \le \ang{30}$, while few complementary triplets shall be placed in nearly polar orbits. This design would allow a complete sky coverage, with the possibility to completely solve the three-dimensional localization of the astrophysical event. With this configuration, all the spacecraft can be pointed in LVLH optimal directions, guaranteeing a positive scientific return and solving the limitations of the pathfinder design. 

The specific values of the scientific performances for the full constellation composed of $N_{SC}$ satellites, along a $\SI{2}{y}$ mission, are: $\SI{\ge15}{GRB/SC}$ for zenith pointing without optimization maneuvers, and $\SI{\ge35}{GRB/SC}$ for LVLH optimal with update period of one week. The lower bound of the specific performances for the full constellation is relative to $N_{SC}=18$, while a smaller constellation typically guarantees better specific values.

\section{Conclusion}

The paper introduced a general framework for the mission analysis and sky visibility evaluation of distributed instruments flying in nano-satellites constellation. The study case was applied to the HERMES constellation, dedicated to the detection and localization of high-energy astrophysical transients.
In particular, the propulsion-less nature of the mission forced an accurate and detailed analysis of the dynamical behavior of the satellites in natural motion. 
The definition of constellation injection strategies to assure the fulfillment of scientific requirements was presented. This analysis was supported by a thorough investigation on the influence of the injection impulse, as well as the most significant gravitational field perturbations. 

In order to overcome the limitations given by the lack of orbital control, resulting in a temporal drift of the constellation, specific pointing strategies for the scientific instruments were discussed. In particular, an optimization routine was proposed to increase the scientific return of the mission, computing the most-performing pointing directions for the coordinated satellites. The available pointing strategies were compared and critically discussed with respect to their influence on astrophysical data return, in terms of accuracy of the localization, coverage of the sky regions and number of observed events.

The resulting constellation baseline entails a quasi-simultaneous injection for multiple spacecraft along three different directions, namely $+h$, $+v$ and $-v$-bar. Such strategy guarantees a relative natural drift that ultimately takes the satellites to the required physical baseline. The pointing strategy is optimized with respect to LVLH pointing directions. 
A comprehensive mission analysis tool, coupled with a high-fidelity dynamical propagator, was developed to assess the scientific performance of the constellation, taking into account all the imposed mission requirements. The available simulated results were used to support the presented discussions, assess the robustness of the proposed design and propose possible extension to a full constellation.

The simulation set-up and the proposed methodology can be easily extended to include any constraint, expressed with the same mathematical formulation. This allows the tool to be implemented for different mission scenarios, involving distributed instrumentation in propulsion-less nano-satellites constellations. In addition, the methodology can be applied to Earth-monitoring and planetary remote sensing constellations, in which the pointing direction and field of view calculation are simply transferred to Earth or planet projections. The proposed analytical and optimization technique can easily include ground related scientific merit parameters, as long as they can be defined with mathematical sets and formulas.

\section*{Conflict of interest}

The authors declare that they have no conflict of interest.

\section*{Acknowledgments}

The authors want to acknowledge the entire HERMES project consortium, composed by the Italian Space Agency (ASI), Italian Institute of Astrophysics (INAF), Politecnico di Milano, Cagliari University, National Institute of Higher Mathematics (INdAM), Skylabs Technology, Deimos Space, Nova Gorica University, Tubingen University, Lor\'{a}nd E\"{o}tv\"{o}s University,  Aalta Lab, C3S Electronics.

The authors want to acknowledge the European Commission for the funding of the HERMES project in the Horizon 2020 framework under Grant agreement ID: 821896.

%% Bibliografia
\bibliographystyle{unsrt}
\bibliography{references}

\begin{thebibliography}{10}

\bibitem{HERMES_SWARM}
Fabrizio Fiore, Luciano Burderi, Tiziana~Di Salvo, Marco Feroci, et~al.
\newblock {HERMES: a swarm of nano-satellites for high energy astrophysics and
  fundamental physics}.
\newblock In {\em Proceedings of Space Telescopes and Instrumentation 2018:
  Ultraviolet to Gamma Ray, 2018, Austin, Texas, United States}, volume 10699.
  International Society for Optics and Photonics, SPIE, 2018.

\bibitem{FUSCHINO2019199}
Fabio Fuschino, Riccardo Campana, Claudio Labanti, Yuri Evangelista, et~al.
\newblock {HERMES}: An ultra-wide band x and gamma-ray transient monitor on
  board a nano-satellite constellation.
\newblock {\em Nuclear Instruments and Methods in Physics Research Section A:
  Accelerators, Spectrometers, Detectors and Associated Equipment}, 936:199 --
  203, 2019.
\newblock Frontier Detectors for Frontier Physics: 14th Pisa Meeting on
  Advanced Detectors.

\bibitem{pardini2001decay}
Carmen Pardini and Luciano Anselmo.
\newblock Decay of spherical satellites.
\newblock {\em The Journal of the astronautical sciences}, 49(2):255--268,
  2001.

\bibitem{PARDINI2006392}
Carmen Pardini, W.~Kent Tobiska, and Luciano Anselmo.
\newblock Analysis of the orbital decay of spherical satellites using different
  solar flux proxies and atmospheric density models.
\newblock {\em Advances in Space Research}, 37(2):392 -- 400, 2006.
\newblock Thermospheric-Ionospheric-Geospheric(TIGER)Symposium.

\bibitem{EGMNasa}
F.G. Lemoine, S.C. Kenyon, J.~K. Factor, R.~G. Trimmer, et~al.
\newblock {The Development of the Joint NASA GSFC and the National Imagery and
  Mapping Agency (NIMA) Geopotential Model EGM96}.
\newblock Technical Report NASA TP - 1998 - 206861, NASA - Goddard Space Flight
  Center, July 1998.

\bibitem{ECSS-E-ST-10-04C}
{European Cooperation for Space Standardization (ECSS)}.
\newblock {Space Engineering -- Space Environment}.
\newblock Standard {E-ST-10-04C}, ECSS, November 2008.

\bibitem{folkner2014planetary}
William~M Folkner, James~G Williams, Dale~H Boggs, Ryan~S Park, and Petr
  Kuchynka.
\newblock {The planetary and lunar ephemerides DE430 and DE431}.
\newblock {\em Interplanetary Network Progress Report}, 196(42):1--81, 2014.

\bibitem{jacchia1970new}
Luigi~Giuseppe Jacchia.
\newblock {New static models of the thermosphere and exosphere with empirical
  temperature profiles}.
\newblock Special Report 313, Smithsonian Institution Astrophysical Observatory
  (SAO), May 1970.

\bibitem{roberts1971analytic}
Charles~E Roberts.
\newblock An analytic model for upper atmosphere densities based upon
  {Jacchia's} 1970 models.
\newblock {\em Celestial Mechanics}, 4(3-4):368--377, 1971.

\bibitem{NOAA}
NOAA.
\newblock {Space Weather Prediction Center}.
\newblock \url{http://https://www.swpc.noaa.gov/}, 2017.
\newblock Accessed: 2017-09-30.

\bibitem{alfriend2009spacecraft}
Kyle Alfriend, Srinivas~Rao Vadali, Pini Gurfil, Jonathan How, et~al.
\newblock {\em Spacecraft formation flying: Dynamics, control and navigation}.
\newblock Elsevier Astrodynamics Series, 2009.

\bibitem{scharf_ff_survey}
D.~P. {Scharf}, F.~Y. {Hadaegh}, and S.~R. {Ploen}.
\newblock A survey of spacecraft formation flying guidance and control (part
  1): guidance.
\newblock In {\em Proceedings of the 2003 American Control Conference, 2003},
  volume~2, pages 1733--1739, June 2003.

\bibitem{meegan2009fermi}
Charles Meegan, Giselher Lichti, P.~N. Bhat, Elisabetta Bissaldi, et~al.
\newblock {The Fermi Gamma-Ray Burst Monitor}.
\newblock {\em The Astrophysical Journal}, 702(1):791--804, August 2009.

\bibitem{gruber2014fermi}
David Gruber, Adam Goldstein, Victoria~Weller von Ahlefeld, P~Narayana Bhat,
  et~al.
\newblock {The Fermi GBM gamma-ray burst spectral catalog: four years of data}.
\newblock {\em The Astrophysical Journal Supplement Series}, 211(1):12, 2014.

\bibitem{colaIAC2019}
Andrea Colagrossi, Jacopo Prinetto, Stefano Silvestrini, Marco Orfano, et~al.
\newblock {Semi-Analytical Approach to Fasten Complex and Flexible Pointing
  Strategies Definition for Nanosatellite Clusters: The HERMES Mission Case
  from Design to Flight}.
\newblock In {\em 70th International Astronautical Congress, 21-25 October
  2019, Washington D.C., USA}, 2019.

\bibitem{PSO}
J.~Kennedy and R.~Eberhart.
\newblock {Particle swarm optimization}.
\newblock In {\em Proceedings of ICNN'95}. International Conference on Neural
  Networks, Perth, WA, Australia, November 1995.

\bibitem{HYBRID}
Richard~H. Byrd, Mary~E. Hribar, and Jorge Nocedal.
\newblock {An Interior Point Algorithm for Large-Scale Nonlinear Programming}.
\newblock {\em SIAM Journal on Optimization}, 9(4):877--900, 1999.

\bibitem{MCNasa}
J.~M. Hanson and B.~B. Beard.
\newblock {Applying Monte Carlo Simulation to Launch Vehicle Design and
  Requirements Analysis}.
\newblock Technical Report NASA TP - 2010 - 216447, NASA - Marshall Space
  Flight Center, September 2010.

\end{thebibliography}

% BackMatters
\clearpage
\section*{Biographies}
\paragraph{Andrea Colagrossi} is a Postdoctoral Research Fellow at the Aerospace Science and Technology Department of Politecnico di Milano. He is responsible for the GNC subsystem and the Mission Analysis of the HERMES project. He is author and co-author of about 20 scientific publications on GNC, small space systems and non-Keplerian dynamics. He has been involved in national and EU/ESA-funded projects for developing innovative techniques for spacecraft GNC.

\paragraph{Jacopo Prinetto} is a PhD candidate at the Aerospace Science and Technology Department of Politecnico di Milano.  He is author and co-author of about 10 scientific publications on low thrust trajectory design and optimization. He has been involved in national and EU/ESA-funded projects in the role of Mission Analyst.

\paragraph{Stefano Silvestrini} is a PhD candidate at the Aerospace Science and Technology Department of Politecnico di Milano.  He is author and co-author of about 15 scientific publications on AI-based GNC, distributed systems and relative dynamics. He has been involved in national and EU/ESA-funded projects for developing innovative techniques for spacecraft GNC in different scenarios, such as missions to small bodies and proximity operations for debris removal.

\paragraph{Mich\`ele Lavagna} is a Full Professor in flight mechanics at the Aerospace Science and Technology Department of Politecnico di Milano. She is the head of the ASTRA team research group. She is author and co-author of more than 300 scientific publications on space engineering topics. She has been the supervisor of 27 PhD researches and 149 Master theses in space engineering.

\end{document}